\documentclass{article}
\usepackage{graphicx}
\usepackage{amsmath}
\usepackage{amsfonts}
\usepackage{amssymb}

\begin{document}

\section{{\protect\large \ {\protect\LARGE  J.L.SYNGE} }}

\qquad\qquad on

\section{\bigskip WHITEHEAD'S PRINCIPLE OF RELATIVITY.}

\bigskip

\subsection{APPENDIX A: Solar Limb-Effect; \ B: Figures}

\subsection{{\protect\large \ Critically edited by A. John Coleman, }}

\subsection{Queen's University, Kingston, ON, Canada.}

\subsection{\bigskip}

\bigskip

\subsection{Dogmatic Opinions and Objective Thoughts of the Editor.
\textit{colemana@post.queensu.ca}}

\bigskip

It was the opinion of Professor Synge in 1951 \ - probably until his death -,
as it is my opinion in 2005 that the evidence for the validity of Einstein's
and \ of Whitehead's theories of Gravitation is roughly of equal value.
\ Neither Synge nor I\ would claim that either is ``Correct'' . \ Certainly,
Whitehead would be the last to do so. \ \ I refer to these theories as GRT
\ and PR.

Since GRT is the dominant faith among current relativists, am I not, in the
words \ of Synge's Introduction, ``attempting to exhume a corpse'' \ by
mentioning Einstein's and Whitehead's Theories in the same breath? Has not our
Sacred College$^{1}$ spoken, dismissing Whitehead with faint praise and a
passsing reference, in 1970?

\bigskip

Both PR and GRT presuppose the validity of \ Einstein's Special theory$\left(
SRT\right)  $. \ C.M. Will, in his \ discussion$^{2}$ of PR, states in
foot-note $\left(  3\right)  $ that, as regards theAdvance of the perihelion,
the Bending of light \ and the Retardation of electromagnetic signals, the two
theories are both within the limits of observation. However, he also claimed ,
by a complicated argument concerning the local gravitational constant, G,
\ that he \ had administered\ the \ \textit{coup de grace } to PR just as, in
my opinion, \textit{QM has done to GRT.}

\bigskip

\subsubsection{(i) John Lighton Synge (1897-1995)}

Synge wrote 13 books and over 200 papers exemplifying the lucidity and sense
of drama that all of us who had the privilege of attending his lectures knew
and admired. Most notable perhaps were his treatises$^{3}$ on Relativity: SRT
in 1956, and \ GRT \ in 1960. Although these are widely regarded as the
\vspace{0.01in}logically clearest available presentations of Einstein's
Theory, a careful reading of his text reveals that he explicitly refuses to
say that he considers GRT \ to be ``correct''.\bigskip

Indeed, in essence, \ on pp. IX/X \ of the Preface of his volume on GRT, \ he
says that while the Principle of Equivalence played a useful role in firing
imagination it makes no sense and that by 1960 should have been abandoned!. It
was because \ Whitehead explicitly rejects this ``Principle'' that in 1922 he
was able to predict the solar limb-effect noted by Halm in 1907.

\bigskip

\qquad In my senior year, 1938/9, as an undergraduaate \ at the University of
\ Toronto I had the good fortune to audit Synge's \ Graduate Course on GRT.
\ At one point, I asked \ ''Sir, what do you make of Whitehead's criticism of
Einstein's theory of measurement?'' \ . \ He brushed me aside curtly, saying
``I never heard of it!'' and went on, leaving me in embarrassed silence. By
1950 , Synge and his former co-author, Alfred Schild, were the centre of a
minor revival of interest in PR. Later, in a Public Lecture \ in Toronto he
stated that he had become disenchanted with GRT, my question came to his mind
and he decided to look seriously at W.'s theory. \ As far as I am aware he was
the first conpetent Mathematical Physicist to do so except, possibly, Eddington.

\bigskip

In 1952 he invited me to give eight lectures to his seminar at theDublin
Advanced Institute to a small group which included Schroedinger - a
challenging task for a 34-year Assistant Professor! \ These dealt with my
thesis, boldy entitled \ ``Relativistic Quantum Mechanics'' , which was based
\ on the late work of Eddington about the Comstants of Nature. \ 

Although I became Synge's \ ``walking companion'' and, later, his friend ,
unfortunately I never discussed Whitehead with him since I did not learn of
his Maryland \ Lectures until after his death. \newline 

\bigskip

I feel that Synge's Lectures are of great importance in the current discussion
\ about Relativity, Quantum Mechanics \ and the Theory of Everything. I
decided to make them more easily available.

\bigskip

\subsubsection{$\left(  ii\right)  $\qquad The Lectures\qquad}

\bigskip

\qquad What follows is a Portable Latex version of\ a hard copy of the
original duplicated notes of Synge's \ three Lectures given at the University
of Maryland in 1951. \ \ Mathematical symbols had been entered by hand so
occasionally were difficult to read. There was one serious error which made
two pages unintelligible. I hope, but cannot guarantee, that my revision is correct.

In three lectures it was impossible for Synge to do justice to the subtlety of
thought about space, time and matter set forth in Whitehead's \ writings \ on
physics, beginning in 1905, with a profound essay for the Aristotelian
Society, to 1922, culminating in\ his three books$^{4},$ The Principles of
Natural Knowledge(PNK), The Concept of Nature(CN), \ and The Principle of Relativity(PR).

Synge gives a clear elegant summary , in modern tensor notation, of the
mathematical arguments \ in seven of the 13 chapters of Part II \ of PR\ .
\ But he makes no allusion to the 4 Chapters of Part I which attempt to
explain what scientists in general and astronomers in particular mean by terms
such as ``metre'' or \ ``second''. Nor did Synge have time to discuss Part III
in which Whitehead explains his use of \ Tensors. For me as an undergraduate ,
Part III was a revelation of the real significance of tensors and their
relation to the theory of Group Representations.

The chapters which Synge chose to discuss provide the immediate material
needed to compare the implications of PR for the tests which were advanced for
GRT and for which \ PR and GR \ are equal but \ he makes only a brief allusion
to PR, Ch. XIV on the Limb Effect. \bigskip

Synge's discussion is correct though condensed. However, I regard it as a
mistake of \ exposition that he replaced ANW's notation, dJ$^{2}$ \ for
the``gravitational potential'' by \ ds$_{g}^{2}.$ \ It could encourage the
casual reader to miss tANW's main criticism of GRT \ stated on p. 83 of PR, to
which I return in in $\left(  iv\right)  $ below.

\bigskip

I am grateful to Cathleen Synge Morawetz - Professor at the Courant Institute,
former President of the American Mathematical Society, Synge's only child and
his Literary Executor - for encouraging me to complete this project in order
to spread knowledge of \ a little-appreciated aspect of her father's work.

I am also greatly indebted to Jordan Bell - a sophomore in the School of
Mathematics of Carleton \ University, Ottawa - \ for turning an imperfect Mss.
into Latex with admirable dispatch and remarkable insight.

\bigskip

\subsubsection{(iii) \ Alfred North Whitehead (1860 -1947)}

\subsubsection{\bigskip}

ANW, as I shall refer to him , is now chiefly renowned as a philosopher whose
famous book$^{5}$ - subtitled \ ``An Essay in Cosmology'' was the focus of the
new movement known as \textit{Process Philosophy. \ }This occurred after 1923
when he became a Professor at Harvard. From 1884 to 1910, his position at
Trinity College, Cambridge required him to prepare students for the Tripos in
the whole range of Pure and Applied Mathematics. Even so, in that period he
created a new branch of mathematics with his \textit{\ Universal Algebra}%
$^{6}.$ Then with his student , Bertrand Russell, wrote the three volume
\textit{Principia Mathematica. }Thoughout this extraordinary creative period
however, it seems that his favourite subject for lectures was Maxwell's
Theory! This confirms my interpretation of Whitehaed as, at heart, a
mathematical physicist attempting to understand the Universe in all its aspects.

\subsubsection{(iv) WILL'S ``coup de grace''}

I have not succeded in understanding Will's \ argument$^{2}$ so, while I have
strong reservations about its validity, I cannot say that it is incorrect. The
paper was put together during a period when Will was finishing Ph. D. studies.
He mentioned that PR is a difficult theory which is not in the category of
\ the ``metric''\ theories to which the \textit{parameterized post-Newtonian(}%
PPN) notation is applicable. \ Yet it is PPN that he uses to dismiss PR. \ I
first became suspicious of Will's argument on reading his Foot- Note$\left(
10\right)  $\ which ascribes to Whitehead the opposite of my understanding of
Whitehead's clear assertions.

\bigskip

The only responsible critical discussion of \ Will's paper$^{2}$ \ of which I
am aware is that of D.R.Fowler $^{7}$\ who claimed that it contains both
physical and philosophical errors.

\qquad$\left(  a\right)  $ Fowler points out that Will assumes \ in his
critique of PR that the mass of our Galaxy is concentrated at the centre of
the Galaxy\ whereas, by smearing the mass uniformily, Will's \ estimate of the
error predicted by PR is reduced by a factor of 100! \ The force of Fowler's
argument is enhanced by the presence of Dark Matter which was unknown when
Fowler or Will wrote.

.\qquad$\left(  b\right)  $ Fowler states further that Synge and, following
him, Will have quite failed to understand the real meaning of Whitehead's
theories. Whitehead remarked in PR that it would be helpful to read the two
previous books$^{4}$ in which he developed his \ concept of space and time and
the role of Lorentz Transformations. Indeed, I found this essential in the
sporadic sallies that I took since 1937, to penetrate the thought of Whitehead
re. Physics. \ Yet I found no reference to CN or PNK in the writings of Synge,
Will or MTW!

If Fowler is correct it would imply that \ PR was banished from serious
conisderation by mainstream Physics because of a paper \ which estimated its
discrepancy as 100 times the actual amount. This suggests the desirability of
\ a careful review of rhe papers of Will and of \ Fowler by an independent
critical study.

\ Whitehead stated that Einstein's theory of measurement \ involves a basic
inconsistency: one does not know the meaning of \ ``distance'' between two
events, specified initially by physically meaningless co-ordinates, until
\ Einstein's Equations have been solved with initial conditions given in terms
of \ ``metre'' and ``second'' which cannot \ be defined until the equations
have been solved! \ I\ have never seen this criticism directly addressed in
the Literature of GRT. Possibly MTW \ thought they did so with their cute
story of the student, the ant and the apple in the opening pages of their
famous treatise. \ But what this story implies is that you can give a meaning
to ``distance'' if you are an ant (or astronomer) who has solved the equation
for geodesics - which you are unable to formulate. \ \textit{Only if you are
born with a God-given intuitive ability to follow geodesics in space-time
\ would you be able to measure anything!}

\textit{\bigskip}

This is such a clear simple inconsistency that I see no way to avoid it. \ \bigskip

\subsubsection{$\left(  v\right)  $ LIMB-EFFECT}

The \ average of frequencies of a line in the solar spectrum at extreme ends
\ of the equator should remove the Doppler Shift. \ One expects this average
to equal the frequency observed at the centre of the Disk for the same line.
The first observation that this is seldom the case, now named the
``limb-effect'', \ is attributed to Halm \ in 1907. \ I have been unable to
discover anything about Halm and would appreciate enlightenment!\bigskip

IF one accepts GRT including the Strong Equivalence Principle$\left(
SSEP\right)  ,$ it follows that spectral shifts due to gravity are
proportional to the frequency of the line .\ \ Thus GRT can throw no light on
the limb-effect.

\ Synge, like Whitehead, does not accept the Equivalence Principle.

``Whitehead's theory of relativity implies that there is an interaction
between the gravitational and electromagnetic fields such that for an atom at
the surface of a star, the Coulomb potential r$^{-1}$between two charges must
be replaced by%

\begin{equation}
\frac{1}{r}(1-\alpha cos^{2}\theta).\ \ \tag{$\left(        1\right)        $}%
\end{equation}
\ \ \ \ \ \ \ \ \ \ \ \ \ \ \ \ \ \qquad\qquad\qquad\ \ \ \ \ \ \ \ 

Here, $\theta$ is the angle between the radius vector\ joining the two
interacting charges and the \ direction of the stellar radius passing through
the $a$tom; $\ \alpha$ is a small constant depending on the strength of the
gravitational field. At the surface of the sun, $\alpha$ = 2.12x 10$^{-6}$ approximately.

The effect of (1) is to perturb the normal energy levels by the term \qquad

\qquad\qquad%
\begin{equation}
\qquad-\alpha\frac{\cos^{2}\theta}{r} \tag{$\left(        2\right)        $}%
\end{equation}
which has axial symmetry about the stellar radius through the atom. An effect
of precisely this symmetry is what is needed to explain the limb-\-effect in
the solar spectrum. One might also hope that this perturbation could account
for the striking differences which have been observed in shifts within the
same solar multiplet.'' \ 

\textit{This long quotation is fron my 1968 paper reproduced in the Appendix
to which the reader is referred. }

I am not aware of any serious \ attempt to explore the consequences of \ PR in
this connection. Yet a huge effort at great expense is devoted to theories
about \ the solar atmosphere which are based on the GRT formula. \ 

It has frequently been remarked that Solar Spectral shifts depend \ not only
on Doppler and pressure\ \ effects but also on the energy levels of \ the
terms of the transition of which the line gives evidence. Miss Adam$^{8}$
noted that shifts within one multiplet frequently differ by amounts of the
order of the basic prediction of Einstein's formula. \ 

\textbf{I therefore propose, as a first step} in unravelling the apparent
complex confusion in observed Solar spectral shifts, that an extensive
precision study be made of shifts within single Multiplets and that an effort
be made to obtain a theoretical understanding of them, using Whitehead's
formula above or otherwise. It is \ conceivable but highly unlikely that the
results of Adam could be explained as a pressure effect so their verification
might well \ sound the death knell for SEP and even for GRT in its present
form. Certainly, such would be the case if they are explicable by the above
formula of Whitehead.

This issue is highly significant since it bears not only \ on our speculations
re \ motions in the solar atmosphere but also on all the cosmological theories
for which spectral shifts are frequently the ONLY\ available relevant data. It
may well happen that the \ $\alpha$, in Whitehead's formula $\left(  2\right)
,$ \ is quite important for small dense stars.

\bigskip

\subsubsection{$\left(  vi\right)  \operatorname{Re}lativity$ and Quantum Mechanics}

\bigskip

It is widely, perhaps even universally , \ recognized that the most important
unsolved problem in physics \ is \ how to reconile these two great theories
\ or, at least to harmonize them. \ In my (dogmatic) opinion no significant
progress has occurred since the work of \ Eddington in the mid 1930's \ His
ideas were dismissed as ``speculative'' or ``philosophical''. In those days
,this latter word was the kiss of death! \ In fact, the ideas of Eddington
\ were dull traditional physics , compared with the wild speculations and
wonderful TV to which String and M-theory have given rise. \ 

Eddington had a very simple idea. \ Assume, as he did , that GRT \ and QM are
\ both correct, \ \ Choose a pproblem which can be \ solved \ by both. \ He
chose a finite uniform universe. \ Equate \ the two solutions leading to a
relation between
\rule{.1in}{.1in}
and $\lambda$ \ This was the basic step which led him to \textit{calculate
\ }seven Constants of Nature accurately to 1 in 10$^{4}.$ Perhaps the stongest
observational support any Theory \ ever had. It was rejected immediately
because it was too ``philsophical'' but the logical conclusion was not
noticed: \textit{\ one or both GRT and QM \ is/are invalid!\ }

\bigskip

In the preface, to the second edition(1958) of his famous reatise on GRT ,
\ \ W. Pauli, one of tthe principal creators of QM, \ wrote

`` .... a clear connexion between the general theory of relativity and quantum
mechanics is not in sight.'' .

In my opinion, \ in the past three decade the hope of finding such a
connection has disappered in a murky fog created by \ the many conflicting
claims ,and coounter-claims and wild speculation of proponents of Gauge,
String and M theories. \ An attractive counter \ opinion to mine can be found
in the 1998 Cambridge \ treatise: STRING THEORY \ , by Joseph Polchinski or in
MATHEMATICAL PHYSICS, 2000, Imperial College Press, the preparatory volume for
the International Mathematical Congess at Imperial College, in 2000.

\bigskip

\qquad I came awayfrom the Comgress , with the feeling that the ablest physics
Graduate Students were being led into wasting their minds on a misdirected
course chasing a chimera. Over beer, two very competent \ Cambridge \ Graduate
students bemoaned \ \ the fact that they had gone to Cambridge hoping to learn
some exciting Physics but were struggling with exotic mathematics with no
connection to Physics.

\bigskip

I taught GRT enthusiastically for eight years but am now disenchanted. \ It is
an all-or-nothing Theory. \ Whitehead is more modest. He proposed PR as a
first step beyond Newtonian Physics taking acccount of SRT. \ This is what QM
needs and may well be comfortable in PR. \ There is need for a follow-up of
Rayner's paper$^{{}}$ exploring the cosmological implications of PR - (cf.
Rayner,1954, Proc. Roy. Soc., London A \textbf{222, 509; }and Synge, Proc.Roy.
Soc., London \textbf{A 226}, 336)J. I assume that\ such has not been pursued
hitherto because of the common obsession with GRT.

\bigskip

\bigskip

\bigskip

\begin{center}
\bigskip NOTES
\end{center}

\begin{enumerate}
\item  MTW: \ C.W. Misner, K.S. Thorne, J.A. Wheeler authors of
\ \ GRAVITATION, W.H. Freeman \&Cpy, 1970.

\item  Clifford M. Will, \ Astrophysical Journal, \textbf{169, }%
141-155\textbf{, }$\left(  1971\right)  $

\item \textbf{\ \ }$\left(  a\right)  $ RELATIVITY -The Special Theory, 1956
\ $\left(  b\right)  $ RELATIVITY -The General Theory,\ 1960
\ \ \ \ \ \ \ \ \ The North Holland Publishing Cpy.

\item \ PNK: An Enquiry Concerning Natural Knowledge. \ CUP, 1919.
\end{enumerate}

CN: The Concept of Nature, CUP, 1920

PR: \ The Principle of Relativity with Applications, CUP, 1922; Reprinted,\ \ \ \ \ \ 

Dover, NY, 2004.

\begin{enumerate}
\item [5]Process and Reality, MacMillan, 1929

\item[6] A treatise on Universal Algebra with Applications,Vol. 1 , Cambridge
University Press, 1898. Apparently, ANW intended to write another volume
devoted to geometry but this plan was interrupted by working with Bertrand
Russell on the Three volumes of Principia Mathematica.

\item[7] Dean R. Fowler, in PROCESS STUDIES, \textbf{4,} 4 $\left(
1974\right)  .$ \ For an insightful discussion of the central ideas of
Whitehead's approach to physics, see also Fowler's article in \ ibid
\textbf{5, }159-174 $\left(  1975\right)  .$

\item[8] Madge Gertude Adam($1912-2001)$ was a \ Solar Astronomer at the
Oxford Observatory from 1935 until her death at 89. Her work on the nature and
magnetic fields of Sunspots gained international recognition. However, she
also took great interest in the Limb Effect to which two of her widely quoted
papers were devoted: Mon. Notices RAS, \textbf{119}, 460-470$\left(
1959\right)  $ ; ibid. \textbf{177}, 687-707$\left(  1976\right)  .$ In these
and several other papers she reports shifts for various multiplets.

\ \ These I found particularly interesting since one would expect that, if the
effect were due to a perturbation proportional to the line frequency, shifts
of different \ lines of a multiplet would be equal . \ This is not observed.
\ I therefore visited the Oxford Observatory to discuss these observations
with Miss Adam. She was paarticularly keen on the observations for iron lines
since she felt that for these her results were quite reliable.

\qquad

This conversation together with the fact, already noted by St John \ in the
twenties, \ that \ the shift depends on the excitation potential associated
\ with the line, redoubled my conviction that observed shifts within a ingle
multiplet could be the key to \ understanding the solar spectrum.
\end{enumerate}

\bigskip

\begin{enumerate}
\item  A. John Coleman, \ Feb.7, \ 2005
\end{enumerate}

\section{\bigskip Synge's Lectures\bigskip}

\subsection{Lecture I: \ BASIC HYPOTHESES}

\subsubsection{\bigskip}

\subsubsection{$1.1$ Introduction}

In 1922 there appeared a book by the late Professor Alfred North Whitehead
entitled \textit{The Principle of Relativity } (Cambridge University Press).
His book contains a clearly formulated theory of gravitation and of
electromagnetism in a gravitational field, and so invites comparison with
Einstein's General Theory of Relativity which had appeared six or seven years earlier.

In attempting such a comparison, one becomes aware of certain psychological
factors. The philosophy of science, being on the whole very little discussed
among active physicists, is naive in the sense that many things are taken for
granted subconsciously. One may believe, for example, that the laws of nature,
including the ones of which we today know nothing, lie locked in blueprints in
a filing cabinet, and that the achievement of the human mind in theoretical
physics is less an act of creation than a successful burglary performed on
this filing cabinet.

If that view is accepted, then once the real blueprint has been exhibted to
the world, any further blueprints for the same structure brought forward at a
later date must be dismissed as forgeries. There is no room for two different
theories covering the same phenomena; if one is right, then the other must be
wrong. That, it may be claimed, is a fair statement of a view widely, if
silently, held. It is a view very difficult to dismiss from our minds, trained
as they are in an old and little-discussed tradition.

Anyone who attempts to describe Whitehead's theory of gravitation has to face
this attitutde on the part of physicists, and if he is a physicist himself his
own thughts are not immune from it. He cannot help feeling that he may be
swimming against the tide, or (to change the metaphor) exhuming a mummy
instead of trying to contribute to the growth of the living body of physics.
To bolster his own confidence, he may be tempted to create an artificial
enthusiasm as a device of propaganda in order to win a hearing for a theory
which has slipped away into a fairly complete oblivion.

There is another difficulty in describing Whitehead's theory. Whitehead was a
philosopher first and a mathematical physicist second. How can one who is not
a philosopher attempt to describe the work of a philosopher? Certainly he
cannot, if the work of the philosopher is philosophy. But if the philosophy is
only a wrapping for physical theory, then the mathematical physicist can take
a savage joy in tearing off this wrapping and showing the hard kernel of
physical theory concealed in it. Indeed there can be little doubt that the
oblivion in which this work of Whitehead lies is due in no small measure to
the effectiveness as insulation of what a physicist can in his ignorance
describe only as the jargon of philosophy. The account of Whitehead's theory
given in these lectures is emphatically one in which the philosophy is
discarded and attention is directed to the essential formulae. And this, it
may be claimed, is as it ought to be. No student of Newtonian mechanics should
be asked to reconstruct it in the form in which it appeared to Newton himself.

The practical physicist who lives among the facts of observation is naturally
impatient of theories except in so far as they assist him in understanding
those facts. He is entitled to ask, of Einstein's theory of gravitation and of
Whitehead's, whether they adequately explain the facts.

In the case of Einstein's theory, the answer might be put like this: For
slowly moving bodies and weak gravitational fields, the Einstein equations
yield a close approximation to Newtonian gravitation. Since the velocities of
celestial bodies are for the most part slow (in comparison with the velocity
of light) and the gravitational fields are weak, we are to expect only minute
deviations from what is predicted by Newtonian mechanics. Three small
deviations in the solar field are predicted and verified by observation within
the limits of the errors of observation; these are rotation of the perihelion
of Mercury, deviation of a light ray passing close to the sun, and spectral
shift towards the red in a gravitational field.

In the case of Whitehead's theory, all the above statements hold. By what
appears to be a rather extraordinary coincidence, there is a formal agreement
between the two theories in the matter of particle orbits and light rays, and
we find ourselves in the strange position of having two theories which both
appear adequate on the basis of observation.

Such a situation is rather unusual in physics. The facts of observation are so
numerous that it would appear easy to find the crucial observation which would
decide in favor of the one theory or the other, or perhaps controvert them
both. The difficulty here lies in the close agreement of both theories with
Newtonian mechanics, with the result that the critical differences are
necessarily very small.

There is however a difference in the character of the two theories. Einstein's
theory is based on a set of non-linear partial differential equations
involving a somewhat ill-defined term, the energy tensor, and specific
applications are exceedingly difficult to work out. The difficulties are not
merely those of mathematical manipulation. There are deeper difficulties in
the sense that one is hardly convinced that certain problem are clearly
formulated; thus in spite of the ingenious manipulations used in handling the
n-body problem,\footnote{Einstein, Infeld and Hoffman, Annals of Mathematics
39, 65 (1938); Einstein and Infeld, Annals of Mathematics 41, 455 (1940) and
Canadian Journal of Mathematics 1, 209 (1949).} the question remains as to
whether these ``bodies'' are singularities in the field, and, if so, what is
meant by a singularity in a Riemannian space with indefinite line-element.

In respect of clarity, Whitehead's theory has much to recommend it, because it
is not a field theory in the technical sense. The problem of n-bodies can be
unequivocally formulated, and the difficulties of solving it are purely
mathematical. Or to take a terrestrial example, it is possible to formulate
mathematically the problem of a fast-moving particle in the presence of heavy
fixed masses on the earth's surface, a relativistic refinement of the problem
of geodetic observations. In fact Whitehead's theory of gravitation has an
applicabilitiy which the General Theory of Relativity lacks by virtue of the
fact that the latter is a non-linear field theory.

This would appear an opportune occasion to venture an expression of
disagreement with what may be called the ``Filing-Cabinet Theory of Scientific
Theories'' (as discussed earlier) and to suggest that scientific theories
might be viewed as statues or models of which there may be several
representing one thing, but definitely man-made and subject to rejection,
destruction, modification and cannibalisation.

What is given in these lectures is a very free translation of Whitehead's
theory. As indicated above, there is no attempt at all to reconstruct
Whitehead's philosophy. I have not thought it necessary to keep his notation.
But I hope that I have not tampered with the essentials of the theory, namely,
the choice of the axiomatic formulae from which everything else follows.

\bigskip

\subsubsection{1.2 The Minkowskian Background}

The first essential thing to observe about Whitehead's theory is that it uses
the space-time of the \textit{Special Theory of Relativit}y, or, more
correctly, the space-time of Minkowski. Mathematically, this means that we
consider a four-dimensional contiuum of events and in it certain privileged
systems of coordinates (x,y,z,t) related to one another by Lorentz
transformations. Given any two adjacent events, their ``separation'' has a
value
\[
dx^{2}+dy^{2}+dz^{2}-c^{2}dt^{2}%
\]
independent of the particular system of coordinates used. These special
coordinates we may call Galileian. We shall not consider the generalisation of
Whitehead's theory to curved space-time.

The constant c occuring above is a universal constant with a value depending
on the chosen units of space and time. We do not say here that c is the
velocity of light, since light is to be regarded as an electromagnetic
phenomenon, and as such will be discussed later.

In this Minkowskian space-time the history of a particle is a curve, as we are
accustomed to think of it in the Special Theory of Relativity. At this point
we come to the essential hypotheses of the Whitehead theory, which may be
presented as answers to the following two questions:

$\left(  a.\right)  $ What is the gravitational field due to a particle?

$\left(  b.\right)  $ How does a particle move in the gravitational field due
to itself and to other particles?

\bigskip

\subsubsection{$1.3$ The field due to a particle or to a set of particles and
the equations of motion}

Let us use the following notation:
\begin{align*}
\text{space coordinates \ }x_{\alpha}(\alpha &  =1,2,3)\text{ for }x,y,z\\
\text{time coordinate \ }x_{4}  &  =ict.
\end{align*}
Throughout we shall give Greek suffixes the range 1,2,3 and Latin suffixes the
range 1,2,3,4, with summation on repetition in both cases. Thus our space-time
coordinates are $x_{r}$ and the fundamental form, invariant under Lorentz
transformation, is
\begin{equation}
dx_{r}dx_{r}. \tag{3.1}\label{3.1}%
\end{equation}

Except for special use of curvilinear coordinates on occasion, tensorial
properties are with respect to Lorentz transformations only. In terms of the
Minkowskian coordinates $x_{r}$, there is no distinction between covariant and
contravariant tensors; we shall in general use subscripts to indicate the
components, rather than superscripts.

For any element $dx_{r}$ we write
\begin{equation}
ds^{2}=\epsilon dx_{r}dx_{r}, \tag{3.2}\label{eq3.2}%
\end{equation}
$\epsilon=1$ \ for space-like directions,

$\epsilon=-1$ \ for time-like directions,

the sign being chosen to make $ds^{2}$ positive. For a time-like element, $ds
$ is the measure of proper time. The unit tangent vector to a world line is

\bigskip%
\begin{equation}
\lambda_{n}=dx_{n}/ds. \tag{3.3}%
\end{equation}

Consider now a time-like world line $L^{\prime}$ (Fig. \ I.1) and a
point-event $P$ which does not lie on it. From $P$ draw the null cone into the
past, cutting $L^{\prime}$ at $P^{\prime}$, say. Let $x_{n}$ be the
coordinates of $P$ and $x_{n}^{\prime}$ those of $P^{\prime}$. Write for
brevity
\begin{equation}
\xi_{n}=x_{n}-x_{n}^{\prime}, \tag{3.4}\label{eq3.4}%
\end{equation}
so that from the null property,%

\begin{equation}
\xi_{n}\xi_{n}=0. \tag{3.5}\label{eq3.5}%
\end{equation}

Draw the tangent to $L^{\prime}$ at $P^{\prime}$ and drop the perpendicular
$PN$ on it. Then, in an obvious notation, for vectors,
\begin{equation}
NP_{n}=\xi_{n}-|P^{\prime}N|\lambda_{n}^{\prime} \tag{3.6}\label{eq3.6}%
\end{equation}
where $\lambda_{n}^{\prime}$ is the unit tangent vector to $L^{\prime}$ at
$P^{\prime}$. From the orthogonality at $N$ we have
\begin{equation}
NP_{n}\lambda_{n}^{\prime}=0, \tag{3.7}\label{eq3.7}%
\end{equation}
and so, since $\lambda_{n}^{\prime}\lambda_{n}^{\prime}=-1$,
\begin{equation}
|P^{\prime}N|=-\xi_{n}\lambda_{n}^{\prime}. \tag{3.8}\label{eq3.8}%
\end{equation}
Then, from (3.6), since $NP_{n}$ is a space-like vector,%

\begin{align}
|NP|^{2}  &  =NP_{n}NP_{n}=[\xi_{n}+(\xi_{p}\lambda_{p}^{\prime})\lambda
_{n}^{\prime}]\tag{3.9}\label{3.9}\\
\times\lbrack\xi_{n}+(\xi_{q}\lambda_{q}^{\prime})\lambda_{n}^{\prime}]  &
=(\xi_{n}\lambda_{n}^{\prime})^{2}.
\end{align}
In view of the sign in (3.8), we have then
\begin{equation}
|NP|=w, \tag{3.10}%
\end{equation}
where
\begin{equation}
w=-\xi_{n}\lambda_{n}^{\prime}. \tag{3.11}%
\end{equation}

If $L^{\prime}$ is straight, and if it is taken as the time axis, then
$\lambda_{\alpha}^{\prime}=0,\lambda_{4}^{\prime}=i$, \ \ and
\begin{equation}
w=-i\xi_{4}=c(t-t^{\prime}). \tag{3.12}\label{eq3.12}%
\end{equation}
Thus if $r$ is the spatial distance of $P$ from $L^{\prime}$, we have

r = c$(t-t^{\prime})=w,$\bigskip

\noindent when we use the fact that$P^{\prime}P$ is a world line corresponding
to velocity $c$. In general, when $L^{\prime}$ is not straight, the invariant
$w$, as defined in (3.11), plays the role of ``distance'' of $P$ from
$L^{\prime}$, and it is in terms of $w$ that Whitehead defines the
gravitational field due to a particle.

Consider now two adjacent events $P$ and \textit{Q}, with coordinates
$x_{n},and\quad x_{n}+dx_{n}$, and a time-like world line $L^{\prime}$,
representing the history of a particle (Fig. I.2).

From $P$ and $Q$ draw null cones into the past, intersecting $L^{\prime}$ at
$P^{\prime},Q^{\prime}$ respectively, with coordinates $x_{n}^{\prime},\quad
$and $x_{n}^{\prime}+dx_{n}^{\prime}$. We proceed to find the element of
proper time $P^{\prime}Q^{\prime}$, which depends only on the events $P,Q$ and
the world line $L^{\prime}$.

Let the equation of $L^{\prime}$ be $x_{n}^{\prime}=x_{n}^{\prime}(s^{\prime
})$, $s^{\prime}$ being proper time on $L^{\prime}$. Then we have
\begin{equation}
(x_{n}^{\prime}-x_{n})(x_{n}^{\prime}-x_{n})=0, \tag{3.14}\label{3.14}%
\end{equation}
and this is an equation for the determination of $s^{\prime}$. On variation of
$x_{n}$, we get, in the notation of (3.4),
\begin{equation}
\xi_{n}(dx_{n}^{\prime}-dx_{n})=0. \tag{3.15}\label{eq3.15}%
\end{equation}
But $dx_{n}^{\prime}=\lambda_{n}^{\prime}ds^{\prime}$, and so
\begin{equation}
wds^{\prime}+\xi_{n}dx_{n}=0; \tag{3.16}\label{eq3.16}%
\end{equation}
hence
\begin{equation}
\partial s^{\prime}/\partial x_{n}=-\xi_{n}/w, \tag{3.17}\label{eq3.17}%
\end{equation}
and from this
\begin{equation}
\frac{\partial x_{m}^{\prime}}{\partial x_{n}}=\frac{dx_{m}^{\prime}%
}{ds^{\prime}}\frac{\partial s^{\prime}}{\partial x_{n}}=-\lambda_{m}^{\prime
}\frac{\xi_{n}}{w}. \tag{3.18}\label{eq3.18}%
\end{equation}
Then, for the element $P^{\prime}Q^{\prime}$ we have
\begin{align}
dx_{r}^{\prime}  &  =-\frac{\partial x_{r}^{\prime}}{\partial x_{s}}%
dx_{s}=w^{-1}\lambda_{r}^{\prime}\xi_{s}dx_{s},\tag{3.19}\label{eq3.19}\\
d{s^{\prime}}^{2}  &  =-dx_{r}^{\prime}dx_{r}^{\prime}=-w^{-2}\lambda
_{r}^{\prime}\xi_{s}dx_{s}\lambda_{r}^{\prime}\lambda_{t}dx_{t}=w^{-2}\xi
_{s}\xi_{t}dx_{s}dx_{t}.\nonumber
\end{align}

With these formal preliminaries cleared away, we come to Whitehead's
definition of the field due to a particle of mass $m$ and world line
$L^{\prime}$: \emph{the field at P is given by the tensor $g_{mn}$, where
$g_{mn}$ is symmetric and such that}
\begin{equation}
g_{mn}dx_{m}dx_{n}=dx_{n}d_{n}+(mk/w)d{s^{\prime}}^{2}, \tag{3.20}%
\label{eq3.20}%
\end{equation}
\emph{for arbitrary $dx_{r}$, where $d{s^{\prime}}^{2}$ is as in (3.19)and $k
$ is a universal constant.}

We recognize in $mk/w$ the analogue of the Newtonian potential.

An equivalent expression of (3.20) is
\begin{align}
g_{mn}  &  =\delta_{mn}+\widetilde{g}_{mn},\tag{3.21}\label{eq3.21}\\
\widetilde{g}_{mn}  &  =(mk/w^{3})\xi_{m}\xi_{n},\nonumber\\
w  &  =-\xi_{n}\lambda_{n}^{\prime}\nonumber
\end{align}

This ``derivation'' of (3.20) or (3.21) is in the old tradition of theoretical
physics, in which one seeks to lure on the reader step by step from the simple
to the complicated. This is interesting historically, because it shows how the
final result was built up in the mind of author, but the bald fact is that
(3.21) is Whitehead's definition of the field due to a particle, and it must
stand or fall, irrespective of the way in which it was built up, on its merits
as a predictor of correct observational results.

The field of \emph{one} particle is then set down as in (3.21). If we have a
set of particles with $\widetilde{g}_{mn}^{\left(  1\right)  }$masses
$m^{(1)},m^{(2)},\ldots$ and world lines $L^{(1)},L^{(2)},\ldots$, the field
due to them all is defined to be
\begin{align}
g_{mn}  &  =\delta_{mn}+\widetilde{g}_{mn}^{\left(  1\right)  }+\widetilde
{g}_{mn}^{\left(  2\right)  }+...,\tag{3.22}\\
\widetilde{g}_{mn}^{\left(  p\right)  }  &  =\frac{m^{\left(  p\right)  }%
k}{w^{\left(  p\right)  3}}\xi_{m}^{\left(  p\right)  }\xi_{n}^{\left(
p\right)  },\nonumber\\
w^{\left(  p\right)  }  &  =-\xi_{n}^{\left(  p\right)  }\lambda_{n}^{\left(
p\right)  }\text{, \ }p=1,2,3...\nonumber
\end{align}

\bigskip the notation being obvious (no summation for $p$).

Thus, as in Einstein's theory, the gravitational field appears as a symmetric
tensor $g_{mn}$. But there is an important difference, because in Einstein's
theory there is no space-time defined topologically which in the $g_{mn}$ are
functions of position, and the $g_{mn}$ seem to have then the responsibility
of determining the topology. The assumption that there is no measure of
separation in space-time except $g_{mn}dx_{m}dx_{n}$ leads to some oddities,
for if a particle is a singularity at which some of the $g_{mn}$ become
infinite, it may become inaccessible through the infinite length of world
lines drawn to it. This sort of thing cannot happen in Whitehead's theory,
because the gravitational field is displayed against a flat 4-space with
Euclidean topology, and an infinity in the $g_{mn}$ causes no embarrassment at
all (anymore than does an infinite potential in Newtonian gravitation).

We have now to supplement the hypothesis (3.22) with a further hypothesis
regarding the motion of particles. For this Whitehead makes a hypothesis very
like that of Einstein: \emph{The world line of a particle satisfies the
variational principle}
\begin{equation}
\delta\int d\bar{s}_{g}=0, \tag{3.23}\label{eq3.23}%
\end{equation}
\emph{where}
\begin{equation}
d\bar{s}_{g}^{2}=-\bar{g}_{mn}dx_{m}dx_{n}, \tag{3.24}\label{eq3.24}%
\end{equation}
\emph{$\bar{g}_{mn}$ being the same as the $g_{mn}$ of (3.22) but with the
omission of the $\widetilde{g}_{mn}$ which is due to the particle itself.} In
fact the self-field is omitted, just as in Newtonian mechanics.

\bigskip

\subsubsection{1.4 The field of a particle at rest}

If a particle exists alone in the universe, then in (3.23) we are to put
\begin{equation}
-d\widetilde{s}_{g}^{2}=\delta_{mn}dx_{m}dx_{n}=dx_{n}dx_{n}, \tag{4.1}%
\label{eq4.1}%
\end{equation}
and from this it follows immediately that the world line is straight.

We seek the field of this particle, and for this field we shall get the
simplest expression by using special axes in space-time with the time-axis
coincident with the world line of the particle. We are then to put in (3.21)
\begin{align}
\lambda_{\alpha}^{\prime}  &  =0,\lambda_{4}^{\prime}=i, \tag{4.2}%
\label{eq4.2}\\
\xi_{\alpha}  &  =x_{\alpha},\xi_{4}=ir,\nonumber\\
w  &  =r\nonumber
\end{align}
(cf. 3.13), where $r$ is the spatial distance of the point of observation $P$
from the fixed particle, for which $x_{\alpha}^{\prime}=0$. Now, $m$ being the
mass of the particle, we have
\begin{equation}
g_{mn}=\delta_{mn}+(mk/r^{3})\xi_{m}\xi_{n}, \tag{4.3}\label{4.3}%
\end{equation}
and so, with Greek suffixes for the range 1,2,3 in accordance with our
convention,
\begin{align}
g_{\mu_{\nu}}  &  =\delta_{\mu\nu}+\frac{mk}{r^{3}}x_{\mu}x_{\nu}\tag{4.4}\\
g_{\mu4}  &  =i\frac{mk}{r^{2}}x_{\mu}\nonumber\\
g_{44}  &  =1-\frac{mk}{r}\nonumber\\
& \nonumber
\end{align}
\ \ 

\bigskip This gives the fundamental form
\begin{align}
\Phi &  =g_{mn}dx_{m}dx_{n}\tag{4.5}\label{eq4.5}\\
&  =g_{\mu\nu}dx_{\mu}dx_{\nu}+2g_{\mu4}dx_{\mu}dx_{4}+g_{44}dx_{4}%
^{2}\nonumber\\
&  =dx_{\mu}dx_{\nu}+(mk/r^{3})(x_{\mu}dx_{\mu})^{2}+2i(mk/r^{2})x_{\mu
}dx_{\mu}dx_{4}+(1-mk/r)dx_{4}^{2}.\nonumber
\end{align}
We now introduce spherical polar coordinates (Fig.I.3), so that\bigskip%
\begin{align}
dx_{\mu}dx_{\mu}  &  =dr^{2}+r^{2}d\Omega,d\Omega\tag{4.7}\label{eq4.7}\\
&  =d\theta^{2}+\sin{\theta d\phi^{2}}^{2},\,x_{\mu}dx_{\mu}=rdr\nonumber
\end{align}
and the form $\Phi$ reads, since $x_{4}=ict$,
\begin{align}
\Phi &  =(1+\frac{mk}{r})dr^{2}+r^{2}d\Omega\tag{4.8}\label{eq4.8}\\
&  -\frac{2mkc}{r}drdt-(1-\frac{mk}{r})c^{2}dt^{2}\nonumber
\end{align}
This may be called the \emph{Whitehead form for the field of a particle at rest.}

Eddington (Nature 113, 192 (1924)) pointed out a remarkable fact: we can
transform the Whitehead form into the Schwarzschild form which occurs in the
General Theory of Relativity! This is certainly surprising in view of the
difference between Whitehead's simple definition of the field as in (3.2$1$)
and the Einstein procedure which involves the solution of non-linear partial
differential equations.

To carry this out, let us make a transformation from $(r,\theta,\phi,t)$ to
$(r,\theta,\phi,\tau)$ of the form
\begin{equation}
ct=c\tau+f(r), \tag{4.9}\label{eq4.9}%
\end{equation}
$f$ \ being a function to be determined later. We get from (4.8)
\begin{equation}
\Phi=(1+\frac{mk}{r})dr^{2}+r^{2}d\Omega-\frac{2mk}{r}dr(cd\tau+f^{\prime
}(r)dr)-(1-\frac{mk}{r})(cd\tau+f^{\prime}(r)dr)^{2}, \label{eq4.10}%
\end{equation}
and this has the following coefficients:
\begin{align}
\text{of }dr^{2}  &  :1+\frac{mk}{r}-\frac{2mk}{r}f^{\prime}(r)\nonumber\\
&  -(1-\frac{mk}{r})[f^{\prime}(r)]^{2},\tag{4.11}\\
\text{of }2cdrd\tau &  :-\frac{mk}{r}-f^{\prime}(r)(1-\frac{mk}{r}%
),\nonumber\\
\text{of }c^{2}d\tau^{2}  &  :-(1-\frac{mk}{r}).\nonumber
\end{align}
If we can make
\begin{equation}
f^{\prime}(r)=-\frac{mk/r}{1-\frac{mk}{r}}=-\frac{mk}{r-mk}, \tag{4.12}%
\label{eq4.12}%
\end{equation}
then the term in $drd\tau$ vanishes, and we get
\begin{align}
\Phi &  =(1-mk/r)^{-1}dr^{2}+r^{2}d\Omega\tag{4.13}\label{eq4.13}\\
&  -(1-mk/r)c^{2}d\tau^{2},\nonumber
\end{align}
which is precisely the Schwarzschild form for the solar field, i.e. the case
of radial symmetry.

But can we satisfy (4.12)? We must remember that in the whole of space $r$ has
the range $0<r<\infty$, and on this account a difficulty appears. The solution
of (4.12) is in fact discontinuous, with infinite value, for $r=mk$, since
(4.12) gives on integration
\begin{align}
f(r)  &  =mk\ln{\frac{mk}{mk-r}}+A\text{ for }r<mk, \tag{ \ \ \ \ \ 4.14}%
\label{eq4.14}\\
f(r)  &  =mk\ln{\frac{mk}{r-mk}}+B\text{ for }r>mk,\nonumber
\end{align}
$A$ and $B$ being constants of integration. Thus we can carry out the
transformation to the Schwarzschild form (4.13) either inside the
``singularity'' $r=mk$ or outside it, but not for both regions at the same
time. Either of (4.14) makes $\tau$ infinite for $r=mk$.

The value $r=mk$ corresponds to the so-called ``Schwarzschild singularity'' of
the General Theory of Relativity, for we shall later identify the universal
constant $k$ with $2G/c^{2}$ where $G$ is the constant of gravitation. We note
that the Whitehead form (4.8) is free from singularity, except of course at
$r=0$, where we expect it. The Schwarzschild singularity makes its appearance
through the attempt to carry out a transformation to get rid of the product
term in $drdt$; it is, from the Whitehead standpoint, an artificial
mathematical singularity which it is .unnecessary to introduce. Nevertheless
the value $r=mk$ is critical in the Whitehead Theory, as we shall see in
Lecture III.

\bigskip

\pagebreak \pagebreak 

\subsection{Lecture II: \ CELESTIAL MECHANICS}

\subsubsection{\bigskip}

\subsubsection{2.1 Equations of orbits}

On account of the transformability of the Whitehead form into the
Schwarzschild form, and on account of the fact that the equations of particle
orbits are essentially the same, viz., in our notation
\begin{equation}
\delta\int d{\overline{s}}_{g}=0, \tag{1.1}%
\end{equation}
it is to be concluded that planetary orbits in the Whitehead field of a fixed
central particle are transforms of the orbits in the Schwarzschild field. That
is in fact true, but it seems worth while to discuss the Whitehead orbits
directly. Note that for the present we are dealing with the fiield of a
massive \emph{particle}; later we shall consider the field of a \textit{sphere
}of finite size.

The differential equations corresponding to (1.1) may be written in the form
\begin{equation}
\frac{d^{2}x_{r}}{d{{\overline{s}}_{g}}^{2}}+
\begin{Bmatrix}
r & \\
m & n
\end{Bmatrix}
\frac{dx_{m}}{d{\overline{s}}_{g}}\frac{dx_{n}}{d{\overline{s}}_{g}}=0,
\tag{$\left(           1.2\right)           $}%
\end{equation}
where $%
\begin{Bmatrix}
r & \\
m & n
\end{Bmatrix}
$ is a Christoffel symbol, but it is easier to use a Lagrangian form. Let us
write
\begin{align}
2L  &  =(1-\frac{mk}{r}){r^{\prime}}^{2}+r^{2}({{\theta}^{\prime}}^{2}+{\sin
}^{2}{\theta}^{{}}{{\phi}^{\prime}}^{2})\tag{1.3}\\
&  -\frac{2mkc}{r}r^{\prime}t^{\prime}-(1-\frac{mk}{r})c^{2}t^{\prime
2}.\nonumber\\
&
\end{align}
where the prime means differentiation with respect to ${\overline{s}}_{g}$ as
parameter. Then (1.2) is equivalent to
\begin{equation}
\frac{d}{d{\overline{s}}_{g}}\frac{\partial L}{\partial r^{\prime}}%
-\frac{\partial L}{\partial r}=0,\text{ etc.}, \tag{1.4}%
\end{equation}
and we have the first integral
\begin{equation}
2L=-1. \tag{1.5}%
\end{equation}

Since $\phi$ and $t$ are ignorable, we have the first integrals
\begin{align}
\frac{\partial L}{\partial\phi^{\prime}}  &  =r^{2}{\sin}^{2}{\theta}%
\phi^{\prime}=\alpha,\tag{1.6}\\
\frac{\partial L}{\partial t^{\prime}}  &  =-mkcr^{\prime}/r-(1-mk/r)c^{2}%
t^{\prime}=-\beta c,\nonumber
\end{align}
where $\alpha$ and $\beta$ are constants, $\alpha$ having the dimensions of a
length, and $\beta$ \textit{zero} dimensions.

The equation
\begin{equation}
\frac{d}{d\overline{s}_{g}}\frac{\partial L}{\partial e^{\prime}}%
-\frac{\partial L}{\partial\theta}=\frac{d}{d\overline{s}_{g}}(r^{2}%
\theta^{\prime})-r^{2}\sin{\theta}\cos{\theta}{\phi^{\prime}}^{2}=0 \tag{1.7}%
\end{equation}
tells us that if we choose axes in space so that initially $\theta=\frac{1}%
{2}\pi,\quad\theta^{\prime}=0$ (as we surely can), then $\theta=\frac{1}{2}%
\pi$ permanently, so that the orbit is plane. We put $\theta=\frac{1}{2}\pi$
and investigate the three equations
\begin{align}
\quad r^{2}\phi^{\prime}  &  =\alpha,\quad mkr^{\prime}/r+(1-mk/r)ct^{\prime
}=\beta,\tag{1.8}\\
-1  &  =(1+mk/r)^{2}){r^{\prime}}^{2}+r^{2}{\phi^{\prime}}^{2}-2mkcr^{\prime
}t/r-(1-mk/r)c^{2}{t^{\prime}}^{2}.\nonumber
\end{align}
Multiplying the last of these by $(1-mk/r)$ and using the second, we get
\begin{equation}
(1-m^{2}k^{2}/r^{2}){r^{\prime}}^{2}+(1-mk/r)r^{2}{\phi^{\prime}}^{2}%
\tag{1.9}\\
-2mkr^{\prime}(\beta-mkr^{\prime}/r)/r-(\beta-mkr^{\prime}/r)^{2}\nonumber\\
=-(1-mk/r),\nonumber
\end{equation}
or
\begin{equation}
{r^{\prime}}^{2}+(1-mk/r)r^{2}{\phi^{\prime}}^{2}-{\beta}^{2}=-(1-mk/r).
\tag{1.10}%
\end{equation}
We now use the first of (1.8) to get an equation homogeneous in $r^{\prime}$
and $\phi^{\prime}$:
\begin{equation}
{r^{\prime}}^{2}+(1-mk/r)r^{2}{\phi^{\prime}}^{2}-(\beta^{2}/\alpha^{2}%
)r^{4}{\phi^{\prime}}^{2}\tag{1.11}\\
+(1/\alpha^{2})r^{4}{\phi^{\prime}}^{2}(1-mk/r)=0.\nonumber
\end{equation}
Then, putting $u=1/r,\quad r^{\prime}/\phi^{\prime}=dr/d\theta=-u^{-2}%
du/d\phi$, and dividing (1.11) by ${\phi^{\prime}}^{2}$, we have
\begin{equation}
u^{-4}(du/d\phi)^{2}+(1-mku)u^{-2}-(\beta^{2}/\alpha^{2})u^{-4}\tag{1.12}\\
+(1/\alpha^{2})(1-mku)u^{-4}=0,\nonumber
\end{equation}
or
\begin{equation}
(du/d\phi)^{2}=f(u), \tag{1.13}%
\end{equation}
where
\begin{equation}
f(u)=\beta^{2}/\alpha^{2}-(1-mku)/\alpha^{2}-u^{2}(1-mku). \tag{1.14}%
\end{equation}
Since this is a cubic in $u$, (1.13) gives $u$ as an elliptic function of
$\phi$, so the orbits are determined.

To compare with Newtonian theory, we differentiate (1.13) and obtain
\begin{equation}
d^{2}u/d{\phi}^{2}=\frac{1}{2}f^{\prime}(u)=\frac{1}{2}mk/\alpha
^{2}-u+(3/2)mku^{2}, \tag{1.15}%
\end{equation}
or
\begin{equation}
d^{2}u/d\phi^{2}+u=\frac{1}{2}mk/\alpha^{2}+(3/2)mku^{2}. \tag{1.16}%
\end{equation}
Now for slow motion and weak fields (the conditions under which we expect
approximation to the Newtonian theory), from (1.3), we have approximately
\[
2L=-c^{2}{t^{\prime}}^{2}=-1,\quad d\overline{s}_{g}=cdt,
\]
and (1.6) gives
\[
r^{2}\phi^{\prime}=r^{2}(d\phi/dt)(dt/d\overline{s}_{g})=c^{-1}r^{2}%
d\phi/dt=\alpha,\;r^{2}d\phi/dt=\alpha c,
\]
so that $\alpha c$ is approximately the angular momentum per unit mass.
Denoting this by $h$ in the usual Newtonian notation, so that
\[
h=r^{2}d\phi/dt,
\]
we may write (1.16) approximately as
\begin{equation}
d^{2}u/d\phi^{2}+u=\frac{1}{2}mkc^{2}/h^{2}+(3/2)mku^{2}, \tag{1.17}%
\end{equation}
in which the last term is relatively very small. This is to e compared with
the Newtonian equation
\begin{equation}
d^{2}u/d\phi^{2}+u=Gm/h^{2}, \tag{1.18}%
\end{equation}
where $G$ is the gravitational constant, and we bring the Whitehead theory
into limiting coincidence with Newtonian theory for slow motion and weak
fields by making the following choice of the universal constant $k$:
\begin{equation}
k=2G/c^{2}. \tag{1.19}%
\end{equation}

\subsubsection{2.2. Rotation of the perihelion}

Since (1.13) and (1.14) are formally the same as the equations which occur in
the Einstein theory, we can investigate the rotation of the perihelion of an
orbit in the same way for both theories. The following seems to be simplest.

The apsides of an orbit (stationary values of $r$) occur for $du/d\phi=0$, or
equivalently $f(u)=0$, where $f$ is as in (1.14). Write
\begin{align}
f(u)  &  =mk(u-u_{1})(u-u_{2})(u-u_{3}),\\
u_{1}+u_{2}+u_{3}  &  =1/(mk)=\frac{1}{2}c^{2}/(mG),\nonumber
\end{align}
this last being a large number. We identify $u_{1}$ and $u_{2}$ with the
reciprocals of the two apsidal distances of the orbit; then the third zero of
$f(u)$, namely $u_{3}$, is large.

By (1.13) the apsidal angle is accurately
\begin{align}
A  &  =\int_{u_{1}}^{u_{2}}[f(u)]^{-1/2}du\tag{2.2}\\
&  =(mk)^{-1/2}\int_{u_{1}}^{u_{2}}[(u_{2}-u)(u_{{}}-u_{1})(u_{3}%
-u)]^{-1/2}du,\nonumber
\end{align}
if we take $u_{1}<u_{2}$. Hence approximately
\begin{equation}
A=(mku_{3})^{-1/2}\int_{u_{1}}^{u_{2}}[(u_{2}-u)(u-u_{1})]^{-1/2}(1+\frac
{u}{2u_{3}})du.\tag{2.3}%
\end{equation}
But
\begin{align}
\int_{u_{1}}^{u_{2}}[(u_{2}-u)(u-u_{1})]^{-1/2}du  &  =\pi,\tag{2.4}\\
\int_{u_{1}}^{u_{2}}[(u_{2}-u)(u-u_{1})]^{-1/2}udu  &  =\frac{1}{2}\pi
(u_{1}+u_{2})\nonumber\\
(mku_{3})^{-1/2}  &  =[1-mk(u_{1}+u_{2})]^{-1/2}=1+\frac{1}{2}mk(u_{1}%
+u_{2}),\text{ approx.},\nonumber\\
1/u_{3}  &  =1/[(mk)^{-1}-u_{1}-u_{2}]=mk,\text{ approx.},\nonumber
\end{align}
and so
\begin{align}
A  &  =[1+\frac{1}{2}mk(u_{1}+u_{2})][\pi+\frac{1}{4}\pi mk(u_{1}%
+u_{2})]\tag{2.5}\\
&  =\pi\lbrack1+\frac{3}{4}mk(u_{1}+u_{2})].\nonumber
\end{align}
Thus the advance of perihelion per orbital revolution is
\begin{equation}
2A-2\pi=\frac{3}{2}\pi mk(u_{1}+u_{2})=3\pi mGc^{-2}(u_{1}+u_{2}),\tag{2.6}%
\end{equation}
and if we denote the semi-axis major of the orbit by $a$ and the eccentricity
by $e$, so that
\begin{align}
1/u_{1}  &  =a(1+e),\quad1/u_{2}=a(1-e),\tag{2.7}\\
u_{1}+u_{2}  &  =2/a(1-e^{2}),\nonumber
\end{align}
then the advance is
\begin{equation}
2A-2\pi=\frac{6\pi mG}{ac^{2}(1-e^{2})}.\tag{2.8}%
\end{equation}
If, finally, we put $mG=4\pi^{2}a^{3}/t^{2}$, where $t$ is the periodic time,
we obtain the now classical formula
\begin{equation}
2A-2\pi=\frac{24\pi^{3}a^{2}}{c^{2}t^{2}(1-e^{2})}.\tag{2.9}%
\end{equation}
This formula for the rotation of perihelion is common to Einstein's General
Theory of Relativity and to Whitehead's theory of gravitation. The only
difference lies in different interpretations of $r$ and $t$, and of course
these differences are significant in the case of planetary orbits.

\subsubsection{2.3 Circular orbits}

Let us discuss circular orbits in the field of a fixed particle. We have to
use not only the first-order equation (1.13), which reads
\begin{equation}
(du/d\phi)^{2}=\beta^{2}/\alpha^{2}-(1/\alpha^{2})(1-mku)-u^{2}(1-mku),
\tag{3.1}%
\end{equation}
but also the second-order equation (1.15) obtained by differentiating it:
\begin{equation}
d^{2}u/d\phi^{2}=\frac{1}{2}mk/\alpha^{2}-u+(3/2)mku^{2}. \tag{3.2}%
\end{equation}
Proceeding without approximation, we note the presence of two constants:

an angular momentum constant
\begin{equation}
\alpha=r^{2}\phi^{\prime}, \tag{3.3}%
\end{equation}
and an energy constant
\begin{equation}
\beta=mkr^{\prime}/r+(1-mk/r)ct^{\prime}, \tag{3.4}%
\end{equation}
where the prime indicates $d/d\overline{s}_{g}$.

Circular orbits for given $\alpha$ are found by putting $d^{2}u/d\phi^{2}=0$
in (3.2); thus the radius $r=1/u$ is to satisfy the quadratic equation
\begin{equation}
r^{2}-2\alpha^{2}r/(mk)+3\alpha^{2}=0, \tag{3.5}%
\end{equation}
or, since $k=2G/c$,
\begin{equation}
r^{2}-\alpha^{2}c^{2}r/(mG)+3\alpha^{2}=0, \tag{3.6}%
\end{equation}
of which the roots are
\begin{equation}
r=\frac{1}{2}[\frac{\alpha^{2}c^{2}}{mG} \pm\sqrt{\frac{\alpha^{4}c^{4}}%
{m^{2}G^{2}}- 12\alpha^{2}}]. \tag{3.7}%
\end{equation}
Thus there exists circular orbits only for values of $\alpha$ satisfying
\begin{equation}
\alpha^{2} \geq12m^{2}G^{2}/c^{4}=3m^{2}k^{2}. \tag{3.8}%
\end{equation}
So there is a lower bound for possible angular momentum in a circular orbit.

By (3.6) we also have
\begin{equation}
\alpha^{2}=\frac{r^{2}}{\frac{rc^{2}}{mG}-3}, \tag{3.9}%
\end{equation}
which tells us that
\begin{equation}
r>3mG/c^{2}=(3/2)mk. \tag{3.10}%
\end{equation}
For any $r$ satisfying this inequality there exists a circular orbit, with
$\alpha^{2}$ given by (3.9). For any $\alpha^{2}$ satisfying (3.8) there exist
two circular orbits except that for any $\alpha^{2}=12m^{2}G^{2}/c^{4} $ there
is only one, with radius $r=6mG/c^{2}$.

The general form of the graph connecting $r$ and $\alpha$ for circular orbits
is shown in Figure II.1.

If we use (3.3) and (3.4), with $r^{\prime}=0$ in the latter, the angular
velocity in a circular orbit is
\begin{equation}
\omega=\frac{d\phi}{dt}=\frac{\alpha}{\beta}\frac{c}{r^{2}}(1-\frac{mk}%
{r})=\frac{\alpha}{\beta}\frac{c}{r^{2}}(1-\frac{2Gm}{c^{2}r}). \tag{3.11}%
\end{equation}
If we put $du/d\phi=0$ in (3.1), we get
\begin{equation}
\frac{\beta^{2}}{\alpha^{2}}=(\frac{1}{\alpha^{2}}+u^{2})(1-mku)=(\frac
{1}{\alpha^{2}}+\frac{1}{r^{2}})(1-\frac{mk}{r}), \tag{3.12}%
\end{equation}
and so by (3.9)
\begin{equation}
\frac{\beta^{2}}{\alpha^{2}}=(\frac{c^{2}}{mGr}-\frac{3}{r^{2}}+\frac{1}%
{r^{2}})(1-\frac{2mG}{c^{2}r})=\frac{c^{2}}{mGr}(1-\frac{2mG}{c^{2}r})^{2}.
\tag{3.13}%
\end{equation}
When we substitute this in (3.11) we find that the angular velocity in a
circular orbit satisfies precisely the Newtonian equation
\begin{equation}
\omega^{2}=mG/r^{3}. \tag{3.14}%
\end{equation}

In Whitehead's relativity, as in Einstein's General Theory, the concept of
force is abandoned in favor of a variational principle as in (1.1). We can
however introduce it in Whitehead's theory in a rather special way in
connection with circular orbits, by taking an analogy with Newtonian
mechanics. For in Newtonian mechanics the force per unit mass on a particle
describing a circular orbit of radius $r$ with angular velocity admits two ``definitions'':

\begin{quote}
(a) by means of angular velocity: $F_{\omega}=\omega^{2}r$; (b) by means of
angular momentum: $F_{h}=h^{2}/r^{3}$.
\end{quote}

Since $h=\omega r^{2}$, we have $F_{\omega}=F_{h}$.

If we carry the analogy into Whitehead's relativity, and use $\alpha_{c}$
instead of $h$, we get the following definitions:

\begin{quote}
(a) by means of angular velocity: $F_{\omega}=\omega^{2}r=mG/r^{2}$ by (3.14);
(b) by means of angular momentum: $F_{\alpha}=\frac{\alpha^{2}c^{2}}{r^{3}}=
\frac{mG}{r^{2}}(1-\frac{3mG}{c^{2}r})^{-1}$.
\end{quote}

Thus $F_{\omega}$ agrees with the Newtonian value, whereas $F_{\alpha}$ tends
to infinity as $r$ tends to $3mG/c^{2}$, which is the radius of the smallest
possible circular orbit.

\subsubsection{2.4. The gravitational field of a sphere}

Suppose we have a finite body at rest. Then, referring to (3.22) and (4.4) of
Lecture I, the field at the point $x_{\mu}$ is given by
\begin{align}
g_{\mu\nu}=\delta_{\mu nu}+k\sum m^{\prime}(x_{\mu}-x^{\prime}_{\mu}) (x_{\nu
}-x^{\prime}_{\nu})/r^{3},\tag{4.1}\\
g_{\mu4}=ik\sum m^{\prime}(x_{\mu}-x^{\prime}_{\mu})/r^{2},\nonumber\\
g_{44}=1-k\sum m^{\prime}/r,\nonumber
\end{align}
where $k=2G/c^{2}$ and the summations are over all the particles forming the
body, $m^{\prime}$ being a typical mass, $x_{\mu}^{\prime}$ a typical
position, and $r^{2}=(x_{\mu}-x_{\mu}^{\prime})(x_{\mu}-x_{\mu}^{\prime})$. We
recall that Greek suffixes have the range 1, 2, 3.

For a continuous distribution of density $\rho(x^{\prime})$, we replace the
summations by integrations:
\begin{align}
g_{\mu\nu}=\delta_{\mu\nu}+k\int\rho(x^{\prime})(x_{\mu}-x_{\mu}^{\prime
})(x_{\nu} -x_{\nu}^{\prime})r^{-3}d\tau^{\prime}\tag{4.2}\\
g_{\mu4}=ik\int\rho(x^{\prime})(x_{\mu}-x_{\mu}^{\prime})r^{-2}d\tau^{\prime
}\nonumber\\
g_{44}=1-k\int\rho(x^{\prime})r^{-1}d\tau^{\prime},\nonumber
\end{align}
$d\tau^{\prime}$ being the element of volume.

Suppose now that the sphere is of uniform density, with radius $a$ and center
at the origin. We have to evaluate the integrals in (4.2), with $\rho$ removed
from the integrands, since it is constant.

From consideration of tensor form it is clear that, if we write
\begin{align}
I_{\mu\nu}  &  =\int(x_{\mu}-x_{\mu}^{\prime})(x_{\nu}-x_{\nu}^{\prime}%
)r^{-3}d\tau^{\prime},\tag{4.3}\\
I_{\mu}  &  =\int(x_{\mu}-x_{\mu}^{\prime})r^{-2}d\tau^{\prime},\nonumber
\end{align}
then these integrals may be expressed in the form
\begin{equation}
I_{\mu\nu}=f(r_{O})x_{\mu}x_{\nu}+g(r_{O})\delta_{\mu\nu},\quad I_{\mu
}=h(r_{O})x_{\mu}, \tag{4.4}%
\end{equation}
where $f,g,h$ are three functions (which we have to evaluate) and $r_{O}$ is
the distance from the center of the sphere to the point $x_{\mu}$, so that
\begin{equation}
r_{O}^{2}=x_{\mu}x_{\mu}. \tag{4.5}%
\end{equation}
To find these three functions $f,g,h$, we take the point $x_{\mu}$ on the
$x_{3}$-axis, so that
\begin{equation}
x_{1}=0,\quad x_{2}=0,\quad x_{3}=r_{O}. \tag{4.6}%
\end{equation}
Then, in terms of spherical polar coordinates $R,\theta,\phi$, the formulae
(4.3) give for $I_{\mu\nu}$
\[
I_{11}=I_{22}=\int_{R=0}^{a}\int_{\theta=0}^{\pi}\int_{\phi=0}^{2\pi}\frac{%
\begin{array}
[c]{c}%
\\
R^{4}{\sin}^{3}{\theta}^{{}}{\cos}^{2}{\theta}^{{}}dRd\theta d\phi
\end{array}
}{(R^{2}+r_{O}^{2}-2Rr_{O}\cos{\theta})^{3/2}}%
\]
the others vanishing, and for $I_{\mu}$
\begin{equation}
I_{3}=\int_{R=0}^{a}\int_{\theta=0}^{\pi}\int_{\phi=0}^{2\pi}\frac
{(r_{O}-R\cos{\theta})R^{2}\sin{\theta} dRd\theta d\phi}{R^{2}+r_{O}%
^{2}-2Rr_{O}\cos{\theta}}, \tag{4.8}%
\end{equation}
the other two vanishing.

Direct integration gives, for exterior points $(r_{O}>a)$,
\begin{align}
I_{11}  &  =I_{22}=(4\pi/15)a^{5}r_{O}^{-3}.\tag{4.9}\\
I_{33}  &  =(4\pi/3)a^{3}r_{O}^{-1}(1-\frac{2}{5}a^{2}r_{O}^{-2})\nonumber\\
&  =(4\pi/3)a^{3}r_{O}^{-1}-(8\pi/15)a^{5}r_{O}^{-3};\nonumber
\end{align}
then by (4.4)
\begin{align}
I_{11}  &  =g(r_{O})=(4\pi/15)a^{5}r_{O}^{-3},\tag{4.10}\\
I_{33}  &  =f(r_{O})r_{O}^{2}g(r_{O})=(4\pi/3)a^{3}r_{O}^{-1}-(8\pi
/15)a^{5}r_{O}^{-3},\nonumber
\end{align}
and so
\begin{align}
f(r_{O})  &  =(4\pi/3)a^{3}r_{O}^{-3}(1-\frac{3}{5}a^{2}r_{O}^{-2}%
),\tag{4.11}\\
g(r_{O})  &  =(4\pi/15)a^{5}r_{O}^{-3}.\nonumber
\end{align}
When these functions are substituted in (4.4), we have the general expression
for $I_{\mu\nu}$ for any exterior point $x_{\mu}$.

Also, from (4.8),
\begin{equation}
I_{3}=-\frac{1}{4}\pi r_{O}^{2}\{(1-a^{2}/r_{O}^{2})^{2}\log{\frac{1+a/r_{O}%
}{1-a/r_{O}}}-2\frac{a}{r_{O}}(1+\frac{a^{2}}{r_{O}^{2}})\}. \tag{4.12}%
\end{equation}
(If the ratio $a/r_{O}$ is small, this complicated expression reduces to the
approximate value $\frac{4\pi}{3}a^{3}r_{O}^{-1}$.) By (4.4) we have
\begin{equation}
I_{3}=h(r_{O})r_{O}, \tag{4.13}%
\end{equation}
and so comparison of (4.12) and (4.13) gives
\begin{align}
h(r_{O})  &  =\frac{1}{4}\pi r_{O}\{2\frac{a}{r_{O}}(1+\frac{a^{2}}{r_{O}^{2}%
})-(1-\frac{a^{2}}{r_{O}^{2}})^{2}\log{\frac{1+a/r_{O}}{1-a/r_{O}}%
}\}\tag{4.14}\\
(  &  =\frac{4\pi}{3}a^{3}r_{O}^{-2}\text{ approximately, for small }%
a/r_{O}).\nonumber
\end{align}
The value of $I_{\mu}$ for any exterior point $x_{\mu}$ is given by
substituting this function in (4.4).

The integral for $g_{44}$ in (4.2) is of course elementary when $\rho$ is a
constant, and thus ,the evaluation of (4.2) is complete, for we have
\begin{equation}
g_{\mu\nu}=\delta_{\mu\nu}+k\rho I_{\mu\nu},\quad g_{\mu4}=ik\rho I_{\mu
},\quad g_{44}=1-km/r_{O}, \tag{4.15}%
\end{equation}
where $m$ is the total mass of the sphere:
\begin{equation}
m=\frac{4\pi}{3}\rho a^{3}. \tag{4.16}%
\end{equation}
The quadratic form for the field is
\begin{align}
\Phi &  =g_{mn}dx_{m}dx_{n}=g_{\mu\nu}dx_{\mu}dx_{\nu}+2ig_{\mu4}dx_{\mu
}cdt-g_{44}c^{2}dt^{2}\tag{4.17}\\
&  =dx_{\mu}dx_{\mu}+k\rho I_{\mu\nu}dx_{\mu}dx_{\nu}-2k\rho I_{\mu}dx_{\mu
}cdt-(1-mk/r_{O})c^{2}dt^{2}\nonumber\\
&  =[1+k\rho g(r_{O})]dx_{\mu}dx_{\mu}+k\rho f(r_{O})(x_{\mu}dx_{\mu}%
)^{2}-2k\rho h(r_{O})x_{\mu}dx_{\mu}cdt-(1-mk/r_{O})c^{2}dt^{2}.\nonumber
\end{align}

Let us now for simplicity write $r$ for $r_{O}$, so that henceforth $r$ is the
distance from the center of the sphere:
\begin{equation}
r^{2}=x_{\mu}x_{\mu}. \tag{4.18}%
\end{equation}
Then
\begin{align}
k\rho g(r)  &  =(4\pi/15)k\rho a^{5}r^{-3}=(1/5)kma^{2}r^{-3},\tag{4.19}\\
k\rho f(r)  &  =(4\pi/3)k\rho a^{3}r^{-3}(1-\frac{3}{5}a^{2}/r^{2}%
)=kmr^{-3}(1-\frac{1}{5}\frac{a^{2}}{r^{2}}),\nonumber\\
k\rho h(r)  &  =\frac{3}{16}km\frac{r}{a^{3}}\{2\frac{a}{r}(1+\frac{a^{2}%
}{r^{2}}-(1-\frac{a^{2}}{r^{2}})\log{\frac{1+a/r}{1-a/r}},\nonumber
\end{align}
or approximately for small $a/r$
\begin{equation}
k\rho h(r)=kmr^{-2}. \tag{4.20}%
\end{equation}
Then, introducing spherical polar coordinates $r,\theta,\phi$ with
$d\Omega=d\theta^{2}+{\sin{\theta}}^{2}d\phi^{2}$, we have from (4.17)
\begin{align}
\Phi &  =[1+k\rho g(r)][dr^{2}+r^{2}d\Omega]+k\rho f(r)r^{2}dr^{2}-2k\rho
h(r)rdrcdt-\tag{4.21}\\
&  (1-mk/r)c^{2}dt\nonumber\\
&  =Adr+Br^{2}d\Omega-2Cdr\cdot cdt-(1-km/r)c^{2}dt^{2}\nonumber
\end{align}
where
\begin{align}
A  &  =1+k\rho\lbrack g(r)+r^{2}f(r)]\tag{4.22}\\
B  &  =1+k\rho g(r),\quad C=k\rho rh(r),\nonumber
\end{align}
or explicitly,
\begin{align}
A  &  =1+\frac{km}{r}(1-\frac{2}{5}\frac{a^{2}}{r^{2}}),\tag{4.23}\\
B  &  =1+\frac{1}{5}km\frac{a^{2}}{r^{3}}\nonumber\\
C  &  =\frac{3}{16}km\frac{r^{2}}{a^{3}}\{2\frac{a}{r}(1+\frac{a^{2}}{r^{2}%
})-(1-\frac{a^{2}}{r^{2}})^{2}\log{\frac{1+a/r}{1-a/r}}\}\nonumber
\end{align}
(for small $a/r$, $C=km/r$ approximately).

We have then in (4.21) the quadratic form expressing \emph{the gravitational
field of a sphere of uniform density, at rest}. It should be compared with
(4.8) of Lecture I, which gives the gravitational field of a \emph{particle}.
We see at once that the field does not depend solely on the total mass ($m$)
of the sphere, as it does in Newtonian gravitation, for the radius $a$ of the
sphere appears in the coefficients $A,B,C$ above. The radius does not,
however, appear in the coefficients of $dt^{2}$ in (4.21), and since this term
is the most important dynamically, the effects due to appearance of the radius
in the \ \ other coefficients will be very small. \ We note that A and B
differ from the corresponding coefficients for the field of a particle by
quantities of the order $kma^{2}r^{-3}$ and this difference is very small
unless the point of observation is close to the surface of the sphere. This is
also true of $C$, which has a curiously complicated form; expansion in powers
of $a/r$ gives
\begin{equation}
C=\frac{km}{r}(1-\frac{1}{5}a^{2}r^{-2}+O(a^{4}r^{-4})). \tag{4.24}%
\end{equation}

Planetary orbits in the field of a finite spherical sun of constant density
can of course be worked out. The method is that of Section 1, and instead of
(1.13) we find, for $u=1/r$, the differential equation
\begin{equation}
(du/d\phi)^{2}=F(u) \tag{4.25}%
\end{equation}
where
\begin{equation}
F(u)=\frac{[\beta^{2}-(1-kmu)]B^{2}/\alpha^{2}-(1-kmu)Bu^{2}}{A(1-kmu)+C^{2}}
\tag{4.26}%
\end{equation}
$\alpha$ and $\beta$ being constants of integration:
\begin{align}
Br^{2}\phi^{\prime}=\alpha\tag{4.27}\\
Cr^{\prime}+(1-km/r)ct^{\prime}=\beta
\end{align}
where the prime indicates differentiation with respect to $\overline{s}_{g}$.

It is possible also to find the gravitational field of a \emph{rotating}
sphere and discuss the orbits of planets in this field. In this case however
retardation plays a part, and it is necessary to have recourse to
approximations which make the work somewhat clumsy. No further account of this
problem will be given here.

\subsubsection{2.5. The two-body problem}

In Whitehead's theory it is possible to set up without difficulty the
equations governing the motion of two particles under the influence of their
mutual gravitational interaction. On account of the retardation involved, they
are differential-difference equations, such as one meets in the
electromagnetic two-body problem.

Let $L$ and $L^{\prime}$ be the two world lines (Figure II.2). We take any
event $P$ on $L$ and any event $P^{\prime}$ on $L^{\prime}$ and draw from them
null cones into the past. Then as in Lecture I,
\begin{align}
(g_{mn})_{P}  &  =\delta_{mn}+km^{\prime}w_{P}^{-3}(\xi_{m}\xi_{n}%
)_{P}\tag{5.1}\\
w_{P}  &  =-(\xi_{n})_{P}(\lambda_{n})_{P_{O}^{\prime}}\nonumber\\
(\xi_{n})_{P}  &  =(x_{n})_{P}-(x_{n})_{P_{O}^{\prime}},\quad(\lambda
_{n})_{P_{O}^{\prime}}=(dx_{n}^{\prime}/ds\prime)_{P_{O}^{\prime}}\nonumber
\end{align}
and the field at $P^{\prime}$ due to $L$ is
\begin{align}
(g_{mn})_{P^{\prime}}  &  =\delta_{mn}+kmw_{P^{\prime}}^{-3}(\xi_{m}\xi
_{n})_{P^{\prime}}\tag{5.2}\\
w_{P^{\prime}}  &  =-(\xi_{n})_{P^{\prime}}(\lambda_{n})_{P_{O}}\nonumber\\
(\xi_{n})_{P^{\prime}}  &  =(x_{n})_{P^{\prime}}-(x_{n})_{P_{O}},\quad
(\lambda_{n})_{P_{O}}=(dx_{n}/ds)_{P_{O}}.\nonumber
\end{align}

The variational principles which define the motion read
\begin{equation}
\delta\int d\overline{s}_{g}=0,\quad\delta\int d{\overline{s}}_{g}^{\prime}=0,
\tag{5.3}%
\end{equation}
and these give the differential equations of motion of the second order:
\begin{align}
\frac{d^{2}x_{r}}{d{\overline{s}}_{g}^{2}}+{\
\begin{Bmatrix}
r & \\
m & n
\end{Bmatrix}
}_{P}\frac{dx_{m}}{d{\overline{s}}_{g}}\frac{dx_{n}}{d{\overline{s}}_{g}}  &
=0,\tag{5.4}\\
\frac{d^{2}x_{r}^{\prime}}{d{{\overline{s}}_{g}^{\prime}}^{2}}+{\
\begin{Bmatrix}
r & \\
m & n
\end{Bmatrix}
}_{P^{\prime}}\frac{dx_{m}^{\prime}}{d{\overline{s}}_{g}^{\prime}}\frac
{dx_{n}^{\prime}}{d{\overline{s}}_{g}^{\prime}}  &  =0.\nonumber
\end{align}
However the calculation of the Christoffel symbol is tedious, and it is easier
to work with Lagrangian forms, writing
\begin{align}
2\Lambda_{P}(x,dx)  &  =(g_{mn}dx_{m}dx_{n})_{P}\tag{5.5}\\
&  =(dx_{n}dx_{n})_{P}+km^{\prime}w_{P}^{-3}(\xi_{n}dx_{n})_{P}^{2}.\nonumber
\end{align}
Let us simplify the notation by dropping the subscript $P$. Thus we write
\begin{equation}
2\Lambda(x,dx)=dx_{n}dx_{n}+km^{\prime}w^{-3}[(x_{n}-x_{n}^{\prime}%
)dx_{n}]^{2}, \tag{5.6}%
\end{equation}
wherein $w$ and $x_{n}^{\prime}$ are to be regarded as functions of
$x_{n}^{\prime}$, obtained by drawing the null cone into the past from $P$ and
using its intersection with the world line $L^{\prime}$. Let a dot indicate
differentiation with respect to $\overline{s}_{g}$. Then we may write (5.6) in
the form
\begin{equation}
2\Lambda(x,\dot{x})=\dot{x_{n}}\dot{x_{n}}+km^{\prime}w^{-3}[(x_{n}%
-x_{n}^{\prime})\cdot{x_{n}}]^{2} \tag{5.7}%
\end{equation}
and with a similar expression for $\Lambda^{\prime}$ we can express the
equations of motion (5.4) in the Lagrangian form
\begin{equation}
\frac{d}{d{\overline{s}}_{g}}\frac{\partial\Lambda}{\partial\dot{x_{n}}}%
-\frac{\partial\Lambda}{\partial x_{n}}=0,\frac{d}{d{\overline{s}}_{g}%
^{\prime}}\frac{\partial\Lambda^{\prime}}{\partial{\dot{x_{n}}}^{\prime}%
}-\frac{\partial\Lambda^{\prime}}{\partial x_{n}^{\prime}}=0. \tag{5.8}%
\end{equation}
These have the first integrals
\begin{equation}
2\Lambda=-1,\quad2\Lambda^{\prime}=-1. \tag{5.9}%
\end{equation}

Rather more explicitly, the first of (5.8) reads
\begin{align}
&  \frac{d}{d{\overline{s}}_{g}}\{\dot{x_{n}}+km^{\prime}w^{-3}(x_{n}%
-x_{n}^{\prime})(x_{m}-x_{m}^{\prime})\dot{x_{m}}\}\hspace{0.9cm}\\
&  -km^{\prime}w^{-3}(x_{m}-x_{m}^{\prime})\dot{x_{m}}\dot{x_{p}}(\delta
_{pn}-\partial x_{p}^{\prime}/\partial x_{n})\\
-km^{\prime}[(x_{m}-x_{m}^{\prime})\dot{x_{m}}]^{2}\frac{\partial}{\partial
x_{n}}(w^{-3})  &  =0.
\end{align}
$\left[  {}\right]  $This equation and its companion corresponding to the
second of (5.8), are not easy to handle, and anything in the nature of a
``general solution'' is out of the question. However, there are two problems
of special simplicity to which attention might be given.

The first of these problems is the quasi-Kepler problem, in which one of the
two particles is much more massive than the other. The orbit of the lighter
particles approximates the Kepler orbit discussed in Sections 1 and 2. It
should be possible to push the approximation further without too great labor,
particularly if the relative velocity is assumed small.

The second problem is the symmetric two-body problem, in which not only are
the masses of the two particles assumed to be equal but a further symmetry is
imposed by suitable initial conditions- viz., as judged by some Galileian
observer, the straight line joining simultaneous positions of the two
particles lies in a fixed plane and has its middle point fixed. It seems
likely that under these conditions the distance between the two particles
steadily increases; it would be interesting to have a solution.

\bigskip\pagebreak 

\bigskip

\bigskip

\subsection{Lecture III: \ \ \ \ \ \ \ \ \ \ \qquad ELECTROMAGNETISM
\ \smallskip}

\bigskip

\subsubsection{ \ \ 3.\ 1. The \ \ field quantities}

As always in the Whitehead theory, the background is the flat space-time of
Minkowski; in it we use the imaginary time-coordinate $x_{4}=ict$. In this
flat space-time there is a gravitational field specified by $g_{mn}$, but we
are not now concerned with the way in which this field is produced-- it is a
given field as far as electromagnetism is concerned.

We shall discuss only electromagnestism in vacuo, with special references to
the propagation of light.

The electromagnetic field is described by two skew-symmetric tensors, or
six-vectors; we shall denote them by
\begin{equation}
F_{mn}=-F_{nm},\quad F^{mn}=-F^{nm}. \tag{1.1}%
\end{equation}
Here the subscript-superscript notation has no particular meaning-- it is
simply a notation to distinguish one set of quantities from another set, the
two sets being for the present quite unrelated.

We now make the following ``physical identification'':
\begin{align}
F_{23}  &  =B_{1},F_{31}=B_{2},F_{12}=B_{3}(\vec{B}=\text{magnetic
induction})\tag{1.2}\\
F_{14}  &  =-iE_{1},F_{24}=-iE_{2},F_{34}=-iE_{3}(\vec{E}=\text{electric field
strength})\nonumber\\
F^{23}  &  =H_{1},F^{31}=H_{2},F^{12}=H_{3}(\vec{H}=\text{magnetic field
strength})\nonumber\\
F^{14}  &  =-iD_{1},F^{24}=-iD_{2},F^{34}=-iD_{3}(\vec{D}=\text{dielectric
displacement})\nonumber
\end{align}
We may also include a 4-vector $J^{+}$, with
\begin{align}
J^{\alpha}  &  =(4\pi/c)j^{\alpha}(\vec{j}=\text{current density})\tag{1.3}\\
J^{4}  &  =4\pi i\rho(\rho=\text{charge density})\nonumber
\end{align}
As previously, Latin suffixes have the range 1, 2, 3, 4 and Greek suffixes the
range 1, 2, 3.

\bigskip

\subsubsection{3.2. Maxwell's equations}

We now accept the following partial differential equations (Maxwell's
Equations):
\begin{equation}
F_{[mn,r]}=0,\qquad F_{,s}^{rs}=J^{r}. \tag{2.1}%
\end{equation}
Here the comma denotes partial differentiation, and
\[
F_{[mn,r]}=F_{mn,r}+F_{nr,m}+F_{rm,n}.
\]
With the identifications (1.2) and (1.3), it is easy to see that these are in
fact the usual Maxwell's equations:
\begin{align}
F_{[23,1]}  &  =0\text{ is equivalent to }div\vec{B}=0;\tag{2.2}\\
F_{[23,4]}  &  =0,\text{ etc. are equivalent to }c^{-1}\frac{\partial\vec{B}%
}{\partial t}+rot\vec{E}=0\nonumber\\
F_{,s}^{\alpha s}  &  =J^{\alpha}\text{ equivalent to }c^{-1}\frac
{\partial\vec{D}}{\partial t}-rot\vec{H}=\frac{-4\pi\vec{j}}{c},\nonumber\\
F_{,s}^{4s}  &  =J^{4}\text{ is equivalent to }div\vec{D}=4\pi\rho.\nonumber
\end{align}

\subsubsection{3. Structural equations}

The number of equations in (2.1) is less than the number of field quantities
(1.1) ($J^{\tau}$ being supposed given). To complete the system, we introduce
what may be called $\mathit{\ strucrural}$ equations as follows, involving the
gravitational field:
\begin{equation}
F_{mn}=g_{mr}g_{ns}F^{rs}. \tag{3.1}%
\end{equation}
These are linear equations expressing $\vec{B}$ and $\vec{E}$ in terms of
$\vec{H}$ and $\vec{D}$. If there is no gravitational field, then
$g_{mn}=\delta_{mn}$, and (3.1) become
\begin{equation}
F_{mn}=F^{mn}, \tag{3.2}%
\end{equation}
or
\begin{equation}
\vec{B}=\vec{H},\quad\vec{D}=\vec{E}, \tag{3.3}%
\end{equation}
as of course is proper.

The gravitational field plays the part of dielectric constant and magnetic
permeability, but more generally, since (3.1) does not separately express
$\vec{D}$ in terms of $\vec{E}$ and $\vec{B}$ in terms of $\vec{H}$.

A word with regard to tensor character. In Whitehead's theory we do not seek
invariance of form with respect to general transformations in space-time, but
only with respect to Lorentz transformations. For the Minkowskian coordinates
$x_{n}$, the Lorentz transformation leaves $dx_{n}dx_{n}$ invariant, and so is
(formally) an orthogonal transformation. This implies that the transformation
laws for covariant and contravariant tensors are the same, and we do not need
the notation of subscripts and superscripts to distinguish them. Thus in
general symbols such as $A_{mn}$ and $A^{mn}$ will denote tensors unrelated to
one another except for whatever connection we may deliberately set up, just as
we set up in (3.1) a connection between $F_{mn}$ and $F^{mn}$.

We now define $g^{mn}$ by
\begin{equation}
g^{mr}g_{ms}=\delta_{s}^{r} \tag{3.4}%
\end{equation}
Then (3.1) may be written equivalently
\begin{equation}
F^{mn}=g^{mr}g^{ns}F_{rs}. \tag{3.5}%
\end{equation}
We may collect our formulae as follows:
\begin{align}
F_{[mn,r]}  &  =0,\quad F_{,s}^{rs}=J^{r},\tag{3.6}\\
F^{mn}  &  =g^{mr}g^{ns}F_{rs}.\nonumber
\end{align}

We find it convenient to introduce another tensor defined by
\begin{equation}
F^{\ast^{rs}}=-\frac{1}{2}i\epsilon^{rsmn}F_{mn}, \tag{3.7}%
\end{equation}
where $\epsilon^{rsmn}$ (we may also write it $\epsilon_{rsmn}$) is the usual
permutation symbol, with value 0 unless the suffixes are distinct, 1 if the
suffixes form the set 1234 or an even permutation thereof, -1 if the suffixes
form an odd permutation of 1234. Explicitly, (3.7) read
\begin{align}
F^{\ast^{23}}  &  =-F_{14},\quad F^{\ast^{31}}=-iF_{24},\quad F^{\ast^{12}%
}=-iF_{34},\tag{3.8}\\
F^{\ast^{14}}  &  =-iF_{23}\quad F^{\ast^{24}}=-iF_{31},\quad F^{\ast^{34}%
}=iF_{12},\nonumber
\end{align}
and they are equivalent to
\begin{equation}
F_{rs}=\frac{1}{2}i\epsilon_{rsmn}F^{\ast^{mn}}. \tag{3.9}%
\end{equation}

We have then identically
\begin{equation}
F_{[23,1]}=F_{23,1}+F_{31,2}+F_{12,3}=-iF_{,r}^{\ast^{4r}} \tag{3.10}%
\end{equation}
with similar results for other suffixes, and hence the equations
$F_{[mn,r]}=0$ of (3.6) are equivalent to $F_{,s}^{\ast^{rs}}=0$. Thus we may
rewrite our basic equations (3.6) in the following form, which shows only
field quantities with superscripts:
\begin{align}
F_{,s}^{\ast^{rs}}  &  =0,F_{,s}^{rs}=J^{r},\tag{3.11}\\
F^{mn}  &  =\frac{1}{2}ig^{mr}g^{ns}\epsilon_{rspq}F^{\ast^{pq}}.\nonumber
\end{align}

\subsubsection{3.4. Geometrical optics}

We consider now the propagation of electromagnetic waves (light waves),
putting $J^{r}=0$ in (3.11). However we shall pass at once to geometrical
optics, using the plans described in the lectures on Hamilton's method. In
other words, we shall deal with the characteristics.

\textbf{Plan A:} We assume a solution of (3.11) of the form
\begin{equation}
F^{mn}=G^{mn}\exp{iS},\quad F^{\ast^{mn}}=H^{mn}\exp{iS}, \tag{4.1}%
\end{equation}
where $G^{mn}$ and $H^{mn}$ are skew-symmetric and ``slowly varying'', whereas
$S$ is ``rapidly varying''. We get then, approximately,
\begin{equation}
F_{,n}^{mn}=iS_{,n}G^{mn}\exp{iS},\quad F_{,n}^{\ast^{mn}}=iS_{,n\quad}%
H^{mn}\exp{iS}, \tag{4.2}%
\end{equation}
and so from (3.11) we derive the system
\begin{align}
G^{mn}S_{,n}  &  =0,\quad H^{mn}S_{,n}=0,\tag{4.3}\\
G^{mn}  &  =\frac{1}{2}ig^{mr}g^{ns}\epsilon_{rspq}H^{pq}.\nonumber
\end{align}
Now $S=const.$ is the history of a phase-wave, and the partial differentiation
equation satisfied by $S$ is to be found by eliminating $G^{mn}$ and $H^{mn}$
from (4.3), an algebraic problem. Before attempting it, let us take up a
second plan.

\textbf{Plan B:} Now we study the propagation of a discontinuity in the field,
such as the propagation of light into darkness. It is true that the ``shock
conditions'' across a discontinuity are not prescribed by the partial
differential equations (3.11), but they are suggested by them, as a limit of
the continuous case.

From (3.11) (in which we put $J^{r}=0$) we have $F_{,s}^{rs}=0$, and so,
integrating through any region of space-time,
\begin{equation}
\int F_{,n}^{mn}dV_{4}=0. \tag{4.4}%
\end{equation}
Hence, by Green's theorem,
\begin{equation}
\int F^{mn}d\sum_{n}=0, \tag{4.5}%
\end{equation}
where $d\sum_{n}$ is a directed element of the 3-space bounding $V_{4}$. Let
us now flatten $V_{4}$ down on a 3-space with equation $S=const.$, and at the
same time allow a discontinuity in $F^{mn}$ to develop in the limit; we get
then from (4.5) as shock condition across $S=const.$
\begin{equation}
\delta F^{mn}S_{,n}=0, \tag{4.6}%
\end{equation}
where $\delta F^{mn}$ represents the jump in $F^{mn}$ on crossing $S=const.$
We have used the fact that $d\sum_{n}$ become in the limit direction ratios of
the normal to $S=const.$, and so proportional to $S_{,n}$.

Thus, from the whole set (3.11), we get the set of shock conditions
\begin{align}
\delta F^{mn}S_{,n}  &  =0,\quad\delta F^{\ast^{mn}}S_{,n}=0,\tag{4.7}\\
\delta F^{mn}  &  =\frac{1}{2}ig^{mr}g^{ns}\epsilon_{rspq}\delta F^{\ast^{pq}%
}.\nonumber
\end{align}
To get the partial differential equation satisfied by $S$, we have to
eliminate the quantities $\delta F^{mn}$ and $\delta F^{\ast^{mn}}$, an
algebraic problem precisely the same as that involved in (4.3).

To emphasize the purely algebraic nature of our problem, let us rewrite it in
new notation: It is required to eliminate $A^{mn}$ and $B^{mn}$ from the
equations
\begin{align}
A^{mn}T_{n}  &  =0,\quad B^{mn}T_{n}=0,\tag{4.8}\\
A^{mn}  &  =\frac{1}{2}ig^{mr}g^{ns}\epsilon_{rspq}B^{pq},\nonumber\\
(A^{mn}  &  =-A^{mn},\quad B^{mn}=-B^{mn}).\nonumber
\end{align}

It may be observed that if these equations are satisfied with non-zero $T_{n}
$, then
\begin{equation}
\det{A^{mn}}=0,\quad\det{B^{mn}}=0. \tag{4.9}%
\end{equation}
These are skew-symmetric determinants of even order, and so they are perfect
squares; in fact
\begin{align}
\det{A^{mn}}  &  =
\begin{vmatrix}
0 & A^{12} & A^{13} & A^{14}\\
A^{21} & 0 & A^{23} & A^{24}\\
A^{31} & A^{32} & 0 & A^{34}\\
A^{41} & A^{42} & A^{43} & 0
\end{vmatrix}
\tag{4.10}\\
&  =(A^{23}A^{14}+A^{31}A^{24}+A^{12}A^{34})^{2},\nonumber
\end{align}
so that (4.9) imply
\begin{align}
A^{23}A^{14}+A^{31}A^{24}+A^{12}A^{34}  &  =0,\tag{4.11}\\
B^{23}B^{14}+B^{31}B^{24}+B^{12}B^{34}  &  =0.\nonumber
\end{align}

We carry out the elimination in (4.8) in two steps. First take the case of
\emph{no gravitational field}. Then $g^{mn}=\delta_{mn}$ and our equations
read
\begin{align}
A^{mn}T_{n}  &  =0,\quad B^{mn}T_{n}=0,\tag{4.12}\\
A^{mn}  &  =\frac{1}{2}i\epsilon_{mnpq}B^{pq}.\nonumber
\end{align}
These equations are invariant under a Lorentz transformation, if $T_{n}$
transforms as a vector and $A^{mn}$ and $B^{mn}$ as tensors. If $T_{n}$ exists
(not zero in all components), it must be space-like, time-like, or null. If it
is space-like, we can choose our frame of reference so that
\begin{equation}
T_{1}\neq0,\text{\quad}T_{2}=T_{3}=T_{4}=0, \tag{4.13}%
\end{equation}
and hence by the first line of (4.12)
\begin{equation}
A^{21}=A^{31}=A^{41}=0,B^{21}=B^{31}=B^{41}=0. \tag{4.14}%
\end{equation}
By the last of (4.12) these imply the vanishing of all the components of
$A^{mn}$ and $B^{mn}$, and so we get no wave. Similarly the case of time-like
$T_{n}$ must be ruled out, and we are left with the sole possibility that
$T_{n}$ is a null vector, so that
\begin{equation}
T_{n}T_{n}=0. \tag{4.15}%
\end{equation}
This, then, is the result of eliminating the $A$'s and $B$'s from (4.12).

We now return to the general case (4.8) and carry out the elimination by a
trick, using the preceding result. Consider the application of a non-singular
linear transformation, not orthogonal. Let us agree that $T_{n}$ is to
transform as a covariant vector, $g_{mn}$ like a covariant tensor, $g^{mn}$
like a contravariant tensor, and $B^{mn}$ also like a contravariant tensor. If
$g$ denotes $\det{g_{mn}}$, then it is known that $g^{\frac{1}{2}}%
\epsilon_{rspq}$ transforms liike a covariant tensor. If we finally decide to
make $g^{\frac{1}{2}}A^{mn}$ transform like a covariant tensor, we see that
(4.8) retain their form under the linear transformation, if they are written
equivalently as
\begin{align}
g^{\frac{1}{2}}A^{mn}T_{n}  &  =0,\quad B^{mn}T_{n}=0,\tag{4.16}\\
g^{\frac{1}{2}}A^{mn}  &  =\frac{1}{2}ig^{mr}g^{ns}g^{\frac{1}{2}}%
\epsilon_{rspq}B^{pq}.\nonumber
\end{align}

Now we know that there exists a linear transformation $L(x \to x^{\prime})$
which makes ${g^{\prime}}^{mn}=\delta_{mn}, g^{\prime}=1$. As a result of $L$,
(4.16) takes on the same form as (4.12), but marked with primes. But we know
that from these primed equations we obtain, as in (4.15),
\begin{equation}
T_{n}^{\prime}T_{n}^{\prime}=0, \tag{4.17}%
\end{equation}
or, equivalently,
\begin{equation}
{g^{\prime}}^{mn}T_{m}^{\prime}T_{n}^{\prime}=0. \tag{4.18}%
\end{equation}
But this is an invariant equation, and so if we now apply the transformation
$L^{-1}(x^{\prime}\to x)$ we get
\begin{equation}
g^{mn}T_{m}T_{n}=0. \tag{4.19}%
\end{equation}
\emph{This, then, is the result of eliminating the $A$'s and $B$'s from
(4.18)}.

Restoring the original notation, we see that the phase-wave of (4.1), of the
shock wave of (4.7), is propagated according to the partial differential
equation
\begin{equation}
g^{mn}S_{,m}S_{,n}=0. \tag{4.20}%
\end{equation}

\subsubsection{3.5. Light rays}

Equation (4.20) may be regarded as the tangential equation of a surface,
$S_{,n}$ being direction ratios of its normal. All surfaces satisfying (4.20)
at an event envelope an elementary cone having its vertex at that event
(Figure III.1). To find the equation of the cone, we denote by $\xi_{n}$ an
elementary generator. Then we have
\begin{equation}
\xi_{n}S_{,n}=0, \tag{5.1}%
\end{equation}
and from the envelope condition,
\begin{equation}
\xi_{n}\delta S_{,n}=0,\text{ \quad provided }g^{mn}S_{,m}\delta S_{,n}=0.
\tag{5.2}%
\end{equation}
Hence
\begin{equation}
\xi_{n}=\theta g^{mn}S_{,m}(\theta\text{ undetermined infinitesimal})
\tag{5.3}%
\end{equation}
and so
\begin{equation}
g_{nr}\xi_{n}\xi_{r}=\theta^{2}g^{mp}S_{,m}S_{,p}=0, \tag{5.4}%
\end{equation}
by (4.20). Thus
\begin{equation}
g_{mn}dx_{m}dx_{n}=0 \tag{5.5}%
\end{equation}
is the equation of the elementary cone.

At each event on a wave-surface $S=const.$ there is an elementary cone which
touches this wave-surface, and the directions of tangency define space-time
curves (bicharacteristics). These curves we call \emph{rays}. It follows then
from (5.3) that for some parameter $u$ a ray satisfies
\begin{equation}
dx_{r}/du=g^{rm}S_{,n}\text{ or }g_{rn}dx_{n}/du=S_{,r}. \tag{5.6}%
\end{equation}
Hence we can derive the differential equations of a light ray. We have
\begin{align}
_{{}}\frac{d^{2}x_{r}}{du^{2}}  &  =g_{,p}^{rn}S_{,n}\frac{dx_{p}}{du}%
+g^{rn}S_{,np}\frac{dx_{p}}{du}\tag{5.7}\\
&  =g_{,p}^{rn}S_{,n}g^{pq}S_{,q}+g^{rn}S_{,np}g^{pq}S_{,q}.\nonumber
\end{align}
But by (4.20)
\begin{equation}
\frac{\partial(g^{p}qS_{,p}S_{,q})}{\partial x_{n}}=2g^{pq}S_{,pn}%
S_{,q}+g_{,n}^{pq}S_{,p}S_{,q}=0, \tag{5.8}%
\end{equation}
and so, omitting details of calculation,
\begin{align}
\frac{d^{2}x_{r}}{d\omega^{2}}  &  =g_{,p}^{rn}g^{pq}S_{,n}S_{,q}-\frac{1}%
{2}g^{rn}g_{,n}^{pq}S_{,p}S_{,q}\tag{5.9}\\
&  =\frac{1}{2}\frac{dx_{s}}{du}\frac{dx_{t}}{du}(g^{rp}g_{st,p}%
-g^{rp}g_{ps,t}-g^{rp}g_{pt,s}).\nonumber
\end{align}
Thus we obtain for light rays the familiar equations
\begin{equation}
\frac{d^{2}x_{r}}{du^{2}}+
\begin{Bmatrix}
r & \\
m & n
\end{Bmatrix}
\frac{dx_{m}}{du}\frac{dx_{n}}{du}=0,g_{mn}\frac{dx_{m}}{du}\frac{dx_{n}}%
{du}=0, \tag{5.10}%
\end{equation}
where $%
\begin{Bmatrix}
r & \\
m & n
\end{Bmatrix}
$ is the usual Christoffel symbol of the second kind.

\subsubsection{3.\ 6. Light rays in the solar field}

We note that the differential equations (5.10) for a light ray agree formally
with those used in the General Theory of Relativity. Further, we have seen
that the Whitehead fundamental form for the field of a massive particle can be
transformed into the Schwarzschild form. It is clear then that the usual
formula will be obtained for the bending of a light ray in the field of a
massive particle, except for the reinterpretation of constants, of no physical
interest on account of the smallness of the effect.

We can however open up new ground by investigating the behavior of light rays
in the field of a \emph{finite sphere}, for which we obtained the form (4.21)
of Lecture II.

For this it is more convenient to use Lagrangian equations equivalent to
(5.10). The Lagrangian is
\begin{align}
2L=A{r^{\prime}}^{2}+Br^{2}({\theta^{\prime}}^{2}+{\sin{\theta}}^{2}%
{\phi^{\prime}}^{2})-2Ccr^{\prime}t^{\prime}- (1-km/r)c^{2}{t^{\prime}}^{2}
\tag{6.1}%
\end{align}
where the prime indicates $\frac{d}{d\lambda}$ ($\lambda$ being a parameter
replacing the $u$ of (5.10)) and $A,B,C$ are functions of $r$ as in (4.23) of
Lecture II.

We know that we may put $\theta=\frac{1}{2}\pi$, and we have the first
integrals
\begin{align}
{\partial L}/{\partial\phi^{\prime}} =Br^{2}\phi^{\prime}=\alpha,\tag{6.2}\\
{\partial L}/{\partial t^{\prime}} = -Ccr^{\prime}-(1-mk/r)c^{2}t^{\prime
}=-\beta c,\nonumber
\end{align}
where $\alpha$ and $\beta$ are constants. Also, as in the last of (5.10), we
have $L=0$, or
\begin{equation}
A{r^{\prime}}^{2}+Br^{2}{\phi^{\prime}}^{2}-2Ccr^{\prime}t^{\prime
}-(1-km/r)c^{2}{t^{\prime}}^{2}=0. \tag{6.3}%
\end{equation}
Then (6.2) gives
\begin{equation}
\{A(1-km/r)+C^{2}\}{r^{\prime}}^{2}+Br^{2}{\phi^{\prime}}^{2}(1-mk/r)=\beta
^{2}, \tag{6.4}%
\end{equation}
or
\begin{equation}
\{A(1-km/r)+C^{2}\}{r^{\prime}}^{2}+Br^{2}{\phi^{\prime}}^{2}(1-km/r)=(\beta
^{2}/\alpha^{2})B^{2}r^{4}{\phi^{\prime}}^{2}, \tag{6.5}%
\end{equation}
so that
\begin{equation}
\{A(1-km/r)+C^{2}\}(dr/d\phi)^{2}+Br^{2}(1-km/r)=(\beta^{2}/\alpha^{2}%
)B^{2}r^{4}, \tag{6.6}%
\end{equation}
or, with $u=1/r$,
\begin{equation}
\{A(1-kmu)+C^{2}\}(du/d\phi)^{2}+Bu^{2}(1-kmu)=(\beta^{2}/\alpha^{2})B^{2}.
\tag{6.7}%
\end{equation}
Thus we have the equation which gives the form of the ray in space:
\begin{align}
(du/d\phi)^{2}=F(u),\tag{6.8}\\
F(u)=\frac{(\beta^{2}/\alpha^{2})B^{2}-Bu^{2}(1-kmu)}{A(1-kmu)+C^{2}}\nonumber
\end{align}
To investigate the bending of a ray passing the sun, we put $u=u_{1}$ at the
point of closest approach; then $F(u_{1})=0$, so that
\begin{equation}
(\beta^{2}/\alpha^{2})B_{1}^{2}-B_{1}u_{1}^{2}(1-kmu_{1})=0 \tag{6.9}%
\end{equation}
where $B_{1}$ is the value of $B$ for $u=u_{1}$. Thus we have
\begin{equation}
\beta^{2}/\alpha^{2}=B_{1}^{-1}u_{1}^{2} (1-kmu_{1}). \tag{6.10}%
\end{equation}
Let us approximate on the basis of small $k$, omitting terms of order $k^{2}$.
Then $C^{2}$ is to be dropped from (6.8), and we have
\begin{equation}
F(u)=A^{-1}(1-kmu)^{-1}\{B^{2}B^{-1}u_{1}^{2}(1-kmu_{1})-Bu^{2}(1-kmu)\}.
\tag{6.11}%
\end{equation}
Now by (4.23) of Lecture II we have approximately
\begin{align}
A=1+kmu(1-2/5a^{2}u^{2}),B=1+1/5kma^{2}u^{5},\tag{6.12}\\
A(1-kmu)=1-2/5kma^{2}u^{3}=B^{-2}.\nonumber
\end{align}
Thus
\begin{align}
F(u)=B^{4}\{B_{1}^{-1}u_{1}^{2}(1-kmu_{1})-B^{-1}u^{2}(1-kmu)\}\tag{6.13}\\
=B^{4}\{u_{1}^{2}-u^{2}-km(u_{1}^{3}-u^{3})-\frac{1}{5}kma^{2}(u_{1}^{5}%
-u^{5})\}\nonumber\\
=u_{1}^{2}-u^{2}+\frac{4}{5}kma^{2}u^{3}(u_{1}^{2}-u^{2})-km(u_{1}^{3}%
-u^{3})-\frac{1}{5}kma^{2}(u_{1}^{5}-u^{5}).\nonumber
\end{align}
Writing $G_{n}(u)=(u_{1}^{n}-u^{n})/(u_{1}^{2}-u^{2})$, we have then
\begin{equation}
[F(u)]^{-\frac{1}{2}}=(u_{1}^{2}-u^{2})^{-\frac{1}{2}}\{1-\frac{2}{5}%
kma^{2}u^{3}+ \frac{1}{3}kmG_{3}+\frac{1}{10}kma^{2}G_{5}\}. \tag{6.14}%
\end{equation}

In passing the sun (with shortest distance $u_{1}^{-1}$ from the sun's
center), a light ray is deviated through an angle
\begin{equation}
\gamma=-\pi+2\int_{u=0}^{u=u_{1}}d\phi=-\pi+2\int_{0}^{u_{1}}[F(u)]^{-1/2}du.
\tag{6.15}%
\end{equation}
Now
\begin{align}
\int_{0}^{u_{1}}(u_{1}^{2}-u^{2})^{-1/2}du  &  =\frac{1}{2}\pi,\tag{6.16}\\
\int_{o}^{u_{1}}(u_{1}^{2}-u^{2})^{-1/2}u^{3}du  &  =\frac{2}{3}u_{1}%
^{3},\nonumber\\
\int_{0}^{u_{1}}(u_{1}^{2}-u^{2})^{-1/2}G_{3}(u)du  &  =2u_{1},\nonumber\\
\int_{0}^{u_{1}}(u_{1}^{2}-u^{2})^{-1/2}G_{5}(u)du  &  =\frac{8}{3}u_{1}%
^{3},\nonumber
\end{align}
and so the deviation is
\begin{equation}
\gamma=-\pi+\pi-\frac{8}{15}kma^{2}u_{1}^{3}+2kmu_{1}+\frac{8}{15}kma^{2}%
u_{1}^{3}, \tag{6.17}%
\end{equation}
these terms representing the contributions from the separate terms of (6.14).
The sun's radius $a$ cancels out, and we arrive at the Einstein formula for
the deviation of a light ray:
\begin{equation}
\gamma=2kmu_{1}=\frac{4Gm}{c^{2}r_{1}}, \tag{6.18}%
\end{equation}
where $r_{1}$, is the shortest distance from the sun's center.

\paragraph{3.7. Red shift in a gravitational field}

The Einstein prediction of a shift toward the red in the spectrum of an atom
radiating in a gravitational field is based on the assumption that the $ds$ of
the General Theory of Relativity measures proper time for the atom, in the
sense that the number of vibrationas of a certain spectral type occurring in
an interval $ds$ is a universal constant, independent of the situation of the
radiating atom. Suppose we make the same assumption in the Whitehead theory
viz. that measures proper time in this sense. Then, for an atom at rest in the
field of a uniform sphere, we have by (4.21) of Lecture II,
\begin{equation}
ds^{2}=(1-km/r)c^{2}dt^{2}, \tag{7.1}%
\end{equation}
where $r$ is the distance of the atom from the center of the sphere and $m$ is
the mass of the sphere. We note that the radius of the sphere does not appear.

On the other hand, for the Schwarzschild form we have; as in (4.13) of Lecture
I,
\begin{equation}
ds^{2}=(1-km/r)c^{2}d^{2} \tag{7.2}%
\end{equation}
By (4.9) of Lecture I we have $dt$, and so (7.1) and (7.2) agree. Thus if we
take the view expressed above, the red shift is the same in Whitehead's theory
as in the General Theory of Relativity.

Actually, Whitehead used a simple model of an atom (Principle of Relativity,
Chap. XIII) and obtained a slightly different result. And, again using a
model, he worked out the limb effect. But it would seem that this, being a
question of the frequency of radiation emitted by an atom, cannot be
effectively handled without a formulation in terms of quantum mechanics.

\subsubsection{3.8. Wave velocity and ray velocity of light}

Since, for any Galileian observer, there exists a well defined spatial
background, we can in Whitehead's theory speak of wave \ velocity and ray
velocity of light much more definitely than we can in the General Theory of
Relativity, where the splitting of space-time into space-like sections is a
very arbitrary procedure.

Consider the history of a phase wave or shock wave with equation $S=0$, $S$
being a function of the space-time coordinates. If we solve for $t$, this
history may be written
\begin{equation}
S=t-\phi(x)=0, \tag{8.1}%
\end{equation}
where $x$ here stands for the three spatial coordinates. This wave advances in
space in the sense of $\phi$ increasing, and the unit normal in the direction
of this advance is
\begin{equation}
n_{\rho}=\phi_{,\rho}(\phi_{,\mu}\phi_{,\mu})^{-1/2}, \tag{8.2}%
\end{equation}
the comma denoting partial differentiation. If $w$ is the wave velocity, an
infinitesimal step $dx_{\rho}$ following the wave for a time $dt$ is
\begin{equation}
dx_{\rho}=wn_{\rho}dt. \tag{8.3}%
\end{equation}
But by (8.1) we have
\begin{equation}
dt=\phi_{,\rho}dx_{\rho}=w\phi_{,\rho}n_{\rho}dt=w(\phi_{,\rho}\phi_{,\rho
})^{1/2}dt, \tag{8.4}%
\end{equation}
and so the wave velocity is
\begin{equation}
w=(\phi_{,\rho}\phi_{,\rho})^{-1/2}. \tag{8.5}%
\end{equation}
The components of normal slowness, in the sense of Hamilton, are
\begin{equation}
\sigma_{\rho}=n_{\rho}/w=\phi_{,\rho}. \tag{8.6}%
\end{equation}

Now we have by (8.1)
\begin{equation}
S_{,\rho}=-\phi_{,\rho},S_{,4}=(ic)^{-1}, \tag{8.7}%
\end{equation}
and we may substitute these values in (4.20), viz. $g^{mn}S_{,m}S_{,n}=0$, to
obtain
\begin{equation}
g^{\mu\nu}\phi_{,\mu}\phi_{,\nu}-2(ic)^{-1}g^{\mu4}\phi_{,\mu}-c^{-2}g^{44}=0.
\tag{8.8}%
\end{equation}
Substitution from (8.6) gives \emph{Hamilton's equation}, satisfied by the
components of normal slowness:
\begin{equation}
\Omega(x,\sigma)=g^{\mu\nu}\sigma_{\mu}\sigma_{\nu}+2ic^{-1}g^{\mu4}%
\sigma_{\mu}- c^{-2}g^{44}=0. \tag{8.9}%
\end{equation}
This is the equation of the reciprocal wave surface at each point of space-time.

The wave velocity $w$ in the spatial direction with direction cosines
$n_{\rho}$ must then satisfy, by (8.6),
\begin{equation}
c^{-2}g^{44}w^{2}-2ic^{-1}g^{\mu4}n_{\mu}w-g^{\mu\nu}n_{\mu}n_{\nu}=0,
\tag{8.10}%
\end{equation}
a quadratic equation with the solutions
\begin{equation}
w/c=(g^{44})^{-1}[g^{\mu4}n_{\mu}\pm\{-(g^{\mu4}n_{\mu})^{2}+g^{44}g^{\mu\nu
}n_{\mu}n\nu\}_{{}}^{1/2}]. \tag{8.11}%
\end{equation}
This equation appears to give two values of $w$ corresponding to each given
normal direction $n_{\rho}$. But we must reject as extraneous any negative
root, $w$ being by definition positive by (8.5).

The ray velocity is also easily found, for by (5.5) we have, for a space-time
displacement along a light ray,
\begin{equation}
g_{\mu\nu} dx_{\mu}dx_{\nu}+ 2ig_{\mu4} dx_{\mu}cdt-g_{44} c^{2} dt^{2}=0.
\tag{8.12}%
\end{equation}
We put $dx_{\rho}=v1_{\rho}dt$, $v$ being the ray velocity and $1_{\rho}$ the
direction cosines of the ray. Then $v$ satisfies the quadratic equation
\begin{equation}
c^{2} g_{44}v^{-2} -2ig_{\mu4}1_{\mu}cv^{-1} - g_{\mu\nu} 1_{\mu}1_{\nu}=0,
\tag{8.13}%
\end{equation}
and so
\begin{equation}
c/v=(g_{44})^{-1}[ig_{\mu4} 1_{\mu}\pm\{-(g_{\mu4}1_{\mu})^{2}+g_{44}g_{\mu
\nu} 1_{\mu}1_{\nu}\}^{1/2}]. \tag{8.14}%
\end{equation}
Here also a negative value is to be rejected as extraneous.

The formulae (8.11) and (8.14) give wave velocity and ray velocity for light
in any gravitational field. let us consider now the gravitational field of a
massive particle at rest, so that, as in (4.4) of Lecture I, we have
\begin{align}
g_{\mu\nu}  &  =\delta_{\mu\nu}+kmr^{-3}x_{\mu}x_{\nu},\quad g_{\mu
4}=ikmr^{-2}x_{\mu},\tag{8.15}\\
g_{44}  &  =1-km/r.\nonumber
\end{align}
it is easy to verify that the conjugate tensor has the following simple form:
\begin{align}
g^{\mu\nu}  &  =\delta_{\mu\nu}-kmr^{-3}x_{\mu}x_{\nu}\quad g^{\mu
4}=-ikmr^{-2}x_{\mu},\tag{8.16}\\
g^{44}  &  =1+km/r.\nonumber
\end{align}

If $\psi_{w}$ is the angle between the radius vector $x_{\rho}$ (drawn from
the massive particle to the point of observation) and the direction $n_{\rho}
$ normal to a wave, then $x_{\rho}n_{\rho}=r\cos{\psi_{w}}$ and (8.11)
combined with (8.16) gives
\begin{equation}
w/c=(1+km/r)^{-1}[kmr^{-1}\cos{\psi_{w}}\pm\{1+kmr^{-1}{\sin{\psi_{w}}}%
^{2}\}^{1/2}], \tag{8.17}%
\end{equation}
and if $\psi_{v}$ is the angle between $x_{\rho}$ and the direction $1_{\rho}
$ of a wave, we have by (8.14) and (8.15)
\begin{equation}
c/v=(1-km/r)^{-1}[-kmr^{-1}\cos{\psi_{v}}\pm\{1-kmr^{-1}{\sin{\psi_{v}}}%
^{2}\}^{1/2}], \tag{8.18}%
\end{equation}
or
\begin{equation}
v/c=(1+km/r)^{-1}[kmr^{-1}\cos{\psi_{v}}\pm\{1-kmr^{-1}{\sin{\psi_{v}}}%
^{2}\}^{1/2}]. \tag{8,19}%
\end{equation}

\bigskip For propagation in a radial direction, the ray direction coincides
with the wave nrmal; sin$\psi_{w}=$ sin$\psi_{v}$ = $0,$ and $\left(  \left(
8.17\right)  \right)  $ is the same as $\left(  8.19\right)  .$ For
\textbf{outward} propagation we have $\psi_{w}^{{}}=$ $\psi_{v}=0$ and so, if
$r\rangle km,$ the \textbf{single }value
\begin{equation}
\frac{w}{c}=\frac{v}{c}=1.
\end{equation}

Whereas, if $r\langle km,$ the \textbf{double }\ value
\begin{equation}
\frac{w}{c}=\frac{v}{c}=-1,or\quad\frac{km/r-1}{km+1}%
\end{equation}%
\[
\]
\ \ \ \ \ \ \ \ \ \
\[
\]

For,inward propagation,we have $\psi_{w}$=$\psi_{\nu}$=$\pi$\ and so the
\textbf{single}value
\begin{equation}
\frac{w}{c}=\frac{v}{c}=\frac{1-km/r}{km+1}%
\end{equation}
\textbf{\ }

if r
$>$%
km,

kbut n\textbf{o value at all, }if \ r
$<$%
km!

\bigskip

Thus the Schwarschild singularity, r = km, shows up as a curious point for the
propagation of light!

\pagebreak \bigskip

\subsection{\bigskip\ \ \ Appendix A}

\subsection{\ \ Limb Effect}

\bigskip

(\textbf{NOTE}. \ For many years, I assumed that the paper below had appeared
in the Proceedings of the International Conference on Relativity and
Gravitation in the USSR which R. M. Erdahl and I attended in 1968 and where it
was delivered and accepted. \ Only in 2003, when my old interest in
Whitehead's theory was reviving, did Prof. V. I. Yukalov inform me \ that the
Proceedings of the Conference were \ never published. AJC ).\bigskip

\bigskip

\bigskip

\begin{center}
WHITEHEAD' S PERTURBATION OF ATOMIC ENERGY LEVELS

A. J. Coleman

Department of Mathematics, Queen's University,

Kingston, Ontario, Canada.

\bigskip
\end{center}

Whitehead's theory of relativity implies that there is an interaction between
the gravitational and electromagnetic fields such that for an atom at the
surface of a star, the Coulomb potential r$^{-1}$between two charges must be
replaced by%

\begin{equation}
\frac{1}{r}(1-\alpha cos^{2}\theta).\ \ \tag{$\left(        1\right)        $}%
\end{equation}
\ \ \ \ \ \ \ \ \ \ \ \ \ \ \ \ \ \qquad\qquad\qquad\ \ \ \ \ \ \ \ 

Here, $\theta$ is the angle between the radius vector\ joining the two
interacting charges and the \ direction of the stellar radius passing through
the $a$tom; $\ \alpha$ is a small constant depending on the strength of the
gravitational field. At the surface of the sun, $\alpha$ = 2.12x 10$^{-6}$ approximately.

The effect of (1) is to perturb the normal energy levels by the small term \qquad

\qquad\qquad%
\begin{equation}
\qquad-\alpha\frac{\cos^{2}\theta}{r} \tag{$\left(        2\right)        $}%
\end{equation}
which has axial symmetry about the stellar radius through the atom. An effect
of precisely this symmetry is what is needed to explain the limb-\-effect in
the solar spectrum. One might also hope that this perturbation could account
for the striking differences which have been observed in shifts within the
same solar multiplet.

The effect of the perturbation (2) acting between all pairs of charge is to
add
\begin{equation}
V^{`}=\Sigma_{i}\alpha\frac{Ze^{2}}{r}\cos^{2}\theta_{i}-\Sigma_{i<l}%
\frac{\alpha e^{2}\cos^{2}\theta_{ij}}{r_{ij}} \tag{$\left(        3\right)
$}%
\end{equation}
to the potential in Schroedinger' s equation. Here, Ze is the charge of the
nucleus; 1%
$<$%
i, j%
$<$%
N , where N is the number of electrons in the atom; $\theta_{i}$ is the angle
between \textbf{r}$_{i}$ and the``vertical'' ; $\theta_{ij}$ is the angle
between \textbf{r}$_{ij}$ and the vertical.

By first-order perturbation theory, the shift in energy of a J,M level is
\begin{equation}
\Delta E_{JM}=\langle JM|V^{`}|JM\rangle\tag{$\left(        4\right)        $}%
\end{equation}

For a Term with total orbital momentum L and spin S ,
\begin{equation}
|JM\rangle=\Sigma_{\mu+\nu=M}\langle L\mu S\nu|JM\rangle\varphi_{L\mu}U_{S\nu}
\tag{$\left(        5\right)        $}%
\end{equation}
where $\langle$L$\mu$ S$\nu$%
$\vert$%
JM$\rangle$ is the vector coupling coefficient, and $\varphi_{L\mu}$ and
$U_{S\nu}$ are, respectively, the appropriate pure orbital and pure spin
functions. Since
\begin{equation}
ccs^{2}\vartheta=\frac{1}{3}+\frac{1}{3}(3\cos^{2}\vartheta-1) \tag{$\left(
6\right)        $}%
\end{equation}
by using the indistinguishability of the electrons, the perturbation (4) can
be expressed in the form%

\begin{equation}
\Delta E_{JM}=<JM||V_{0}|JM>+<JM|V_{2}|JM>, \tag{$\left(        7\right)
$}%
\end{equation}
\newline where
\begin{equation}
V_{0}=\frac{\alpha}{3}N\left(  \frac{Ze^{2}}{r_{1}}-\frac{N-1}{r_{12}}%
e^{2}\right)  ,\quad and \tag{$\left(        8\right)        $}%
\end{equation}

\bigskip%

\[
V_{2}=\frac{\alpha}{3}Ne^{2}\left[  \frac{Z}{r_{1}}\left(  3\cos^{2}%
\vartheta_{1}-1\right)  -\frac{N-1}{2r_{12}}\left(  3\cos^{2}\vartheta
_{12}-1\right)  \right]
\]

$\qquad\qquad\qquad\qquad\qquad\qquad\qquad\qquad\qquad\qquad\qquad
\qquad\qquad\left(  9\right)  $

\qquad\qquad\qquad\qquad\qquad

The advantage of this decomposition is that with respect to simultaneous
rotation of all electrons about the nucleus, V$_{0}$ and V$_{2}$ belong to
D$_{0}$ and D$_{2}$ \ \ representation of the rotation group, respectively.

An application of the Wigner-Eckhart theorem leads to the conclusion that
\begin{equation}
\Delta E_{JM}=A_{L}+\frac{3M^{2}-J\left(  J-1\right)  }{J\left(  2J-1\right)
}B_{J} \tag{$\left(        10\right)        $}%
\end{equation}

where
\begin{equation}
A_{L}=\langle\varphi_{LL}|V_{0}|\varphi_{LL}\rangle,\quad B_{J}=\langle
JJ|V_{2}|JJ\rangle. \tag{$\left(        11\right)        $}%
\end{equation}

By employing (5) and the theory of vector-coupling coefficients, a rather
tedious calculation results in the formula%

\begin{equation}
B_{J}=\Sigma_{\mu+\nu=J}\left|  \langle L\mu S\nu|JJ\rangle\right|  ^{2}%
\frac{3\mu^{2}-L\left(  L+1\right)  }{L\left(  2L-1\right)  }B_{L}=
\tag{$\left(       12\right)        $}%
\end{equation}

\qquad\bigskip$\{1+\frac{3(J-L-S)(J+S-L+1)[(J+L-S+1)(J+L+S+2).-2J-3]}{L\left(
2L-1\right)  \left(  2J+2\right)  \left(  2J+3\right)  }\}B_{L}$

\bigskip

where
\begin{equation}
B_{L}=\langle\varphi_{LL}|V_{2}|\varphi_{LL}\rangle. \tag{$\left(
13\right)  $}%
\end{equation}

It follows from the Virial Theorem that%

\begin{equation}
A_{L}=-2\alpha E_{L} \tag{$\left(        14\right)        $}%
\end{equation}

where E$_{:L}$ is the total energy of the state $\varphi_{LL}$ which is given
with sufficient accuracy for the present purposes by the mean observed energy
of the Term. Thus for a Fraunhofer line, the V$_{0}$ tern gives rise to a
red-shift which is proportional to the wave\-length of the line and equal to
2/3 of that predic$t$ed by Einstein.

We have thus reduced the problem of calculating the Whitehead shift in the
levels of a Term to that of evaluating the one constant B$_{L}$. Consequently,
the shifts in the lines of a multiplet depend on two constants B$^{i}$ ,
B$^{f}$ associated with initial and final levels.

Dr. R M. Erdahl has suggested that in attempting to check this theory against
observations we should treat B$^{i}$ and B$^{f}$\ as phenomenological
constants. In certain cases B$_{J}$ = 0 , so that for these the predictions
are particularly simple. For example, from (13) it follows immediately that
B$_{L}$ = 0 if \ L = 0, that is for $an$ S-term. But it follows from (12) that
B$_{J}$ also vanishes for states such as $^{4}$P$_{1/2}$ , $^{6}$%
D$_{1^{\prime}2},$ ... $^{10}$F$_{11/2}$ and many others. \ It may also be
worth looking at Terns for which B$_{J}$ is small. \ 

To test the usefulness of Whitehead's perturbation in explaining the actual
complex observations of shifts in the solar s$p$ectrum, it would be
particularly valuable to have reliable measurements for the absolute shifts at
various points in the solar disc for all lines of a multiplet and especially
for multiplets which include one or more transitions between energy levels
with symmetry type appearing in the list \ described above.

In addition to possible perturbation of energy levels by a gravo-electric
interaction, the Fraunhofer lines are undoubtedly shifted by Doppler and
pressure effects. To this must be added the classic Einstein shift which has
been confirmed by the Pound-Rebka experiment and which follows from Newton's
theory and the conservation of energy. The Einstein and Doppler shifts are
proportional to the wave-length of the line and by themselves certainly cannot
explain the observed shifts in the solar spectrum.

If Whitehead's perturbation combined with reason\-able assumptions about
pressure shift is unable to ex\-plain the observations, all is not lost. If
the astron\-omers can obtain reliable observations, especially at the limb$, $
of a large number of multiplets of diverse symmetry, it should be possible,
using the techniques of the present paper, to obtain a good approximation for
a perturbation of atomic energy levels which would explain the observations by
employing a multipole analysi$s$.

Since in the solar spectrum, the observed deviations from Einstein's predicted
shift are as large or larger than his prediction, it is clearly of great
interest to establish the source of this deviation in order to be able to
interpret spectral shifts from other stars with any confidence.

\bigskip

August 16, 1968.

\clearpage

\section{Appendix B: Figures}

\bigskip\bigskip\begin{center}Figure I.1\end{center}\bigskip\bigskip

\begin{figure}[h]
\begin{center}
\includegraphics[height=4.0in]
{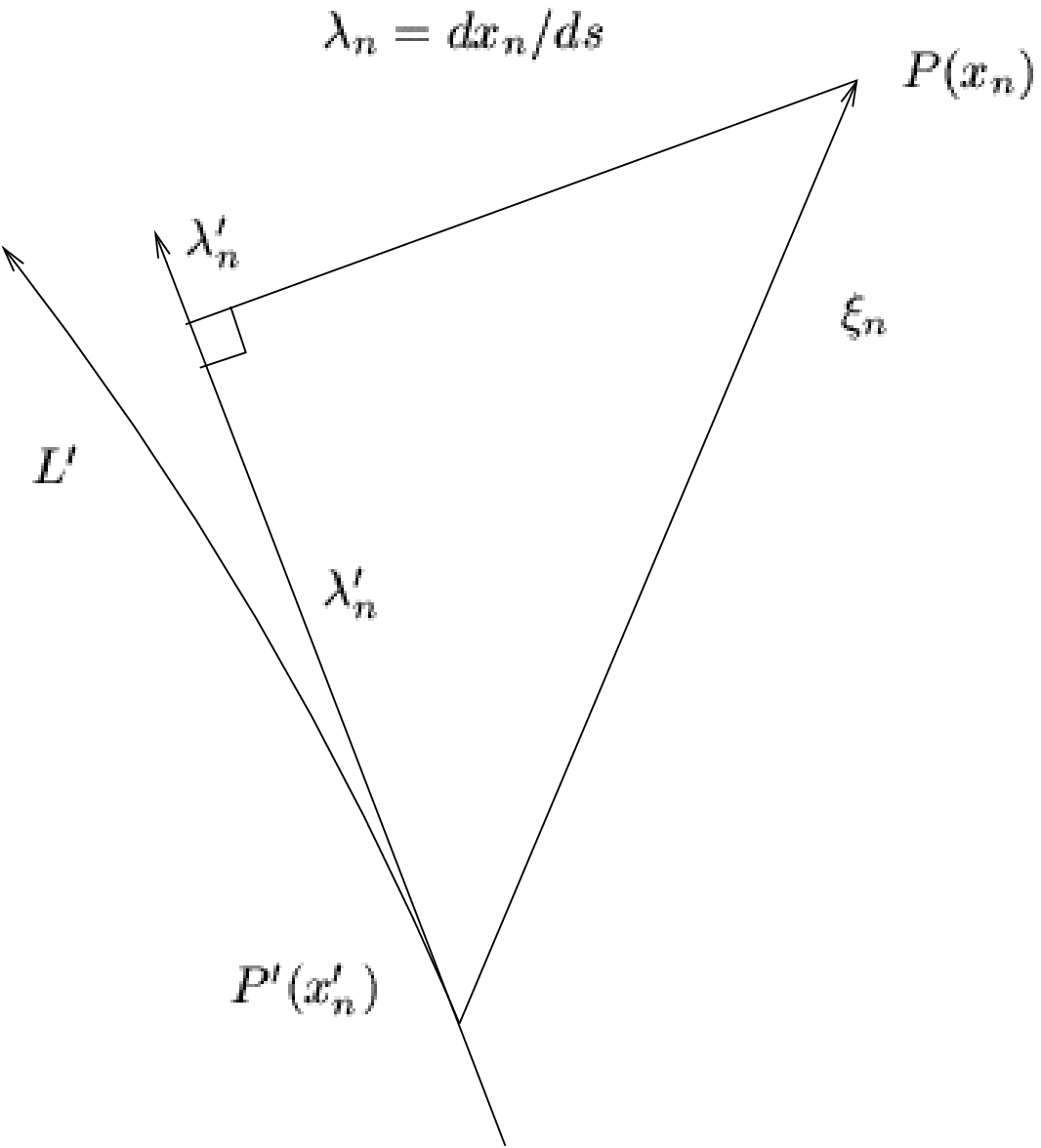}
\end{center}
\end{figure}

\clearpage

\section{Appendix B: Figures}

\bigskip\bigskip\begin{center}Figure I.2\end{center}\bigskip\bigskip

\begin{figure}[h]
\begin{center}
\includegraphics[height=4.0in]
{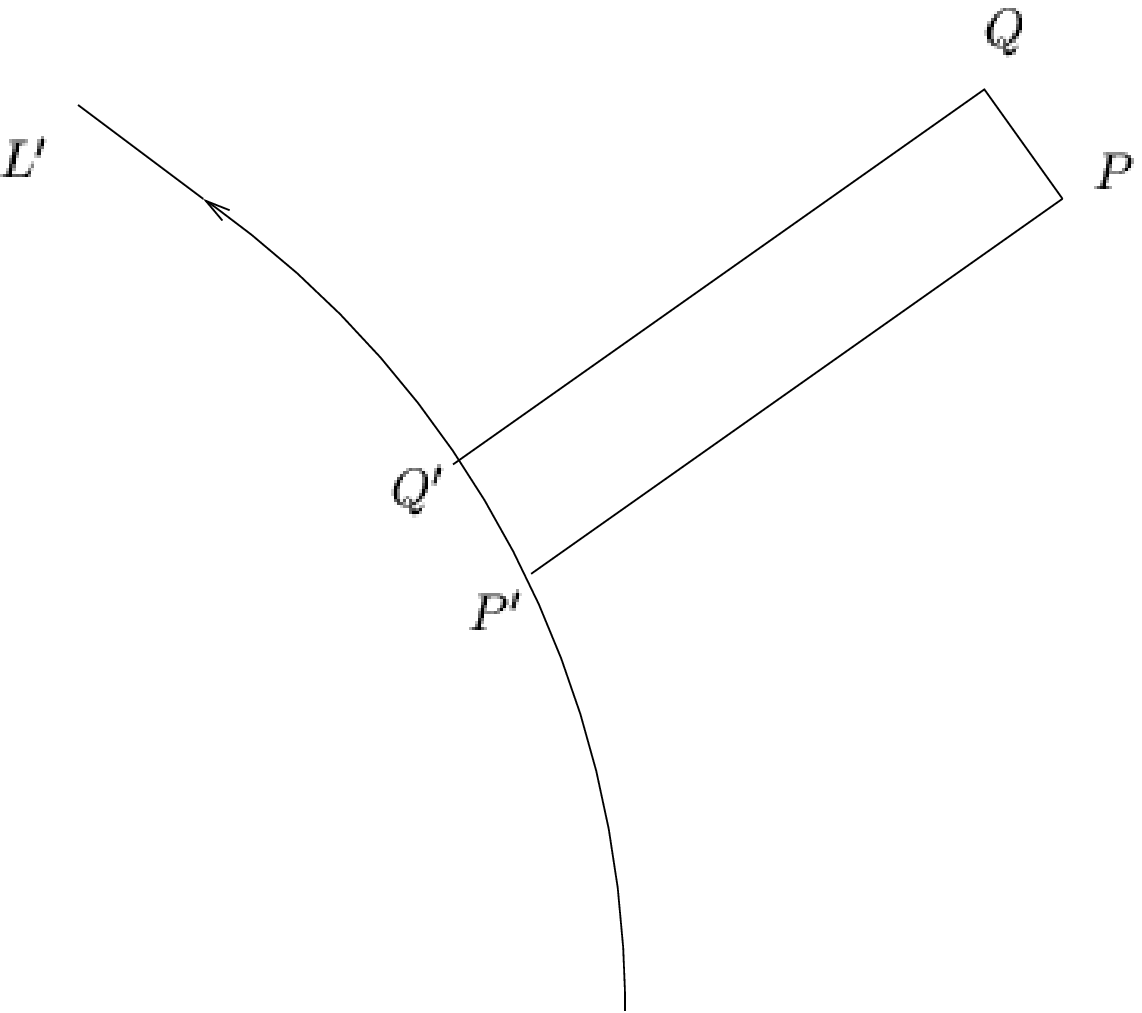}
\end{center}
\end{figure}

\clearpage 

\section{Appendix B: Figures}

\bigskip\bigskip\begin{center}Figure I.3\end{center}\bigskip\bigskip

\begin{figure}[h]
\begin{center}
\includegraphics[height=3.5in]{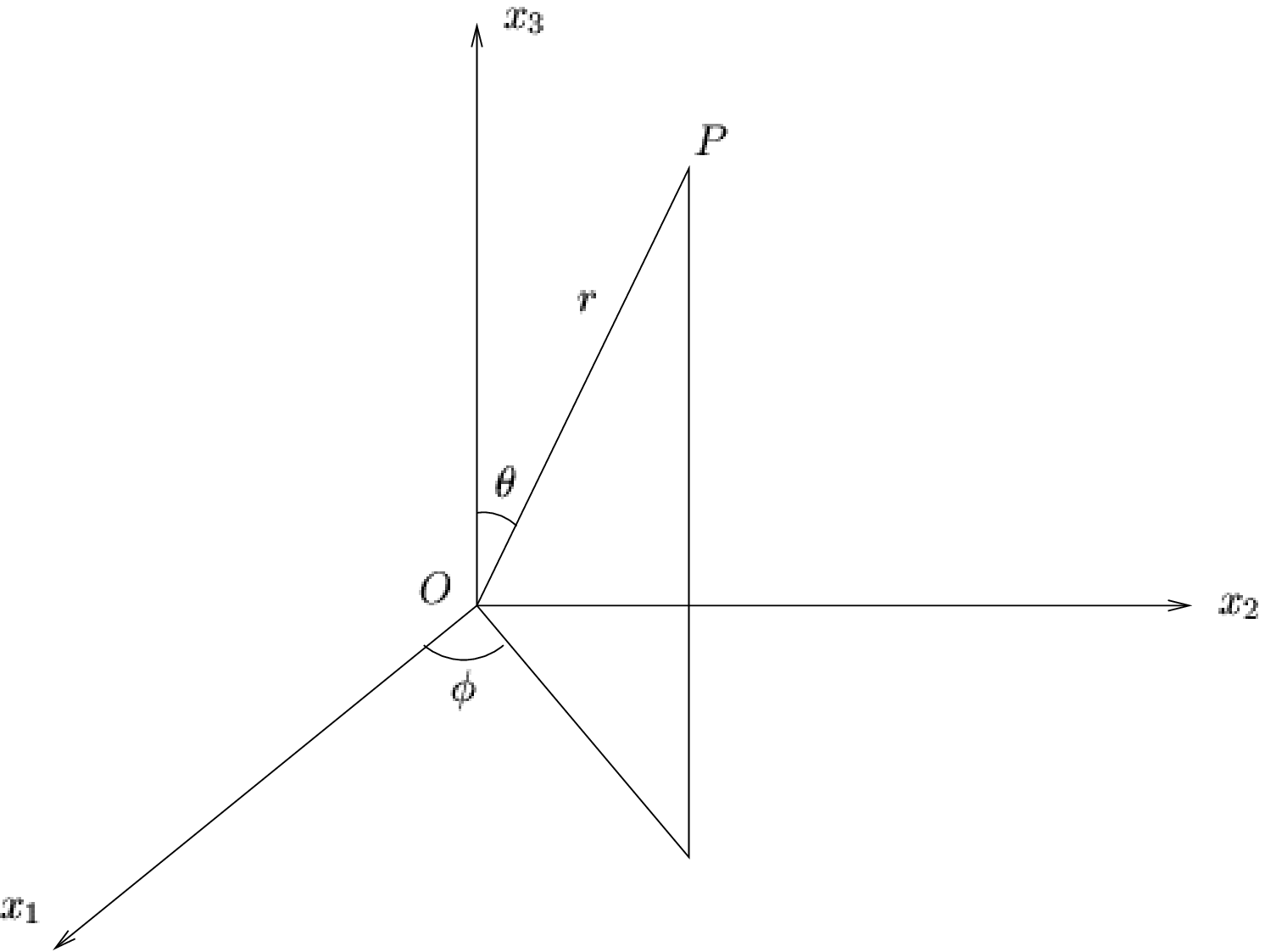}
\end{center}
\end{figure}

\clearpage 

\section{Appendix B: Figures}

\bigskip\bigskip\begin{center}Figure II.1\end{center}\bigskip\bigskip

\begin{figure}[h]
\begin{center}
\includegraphics[width=5.0in]{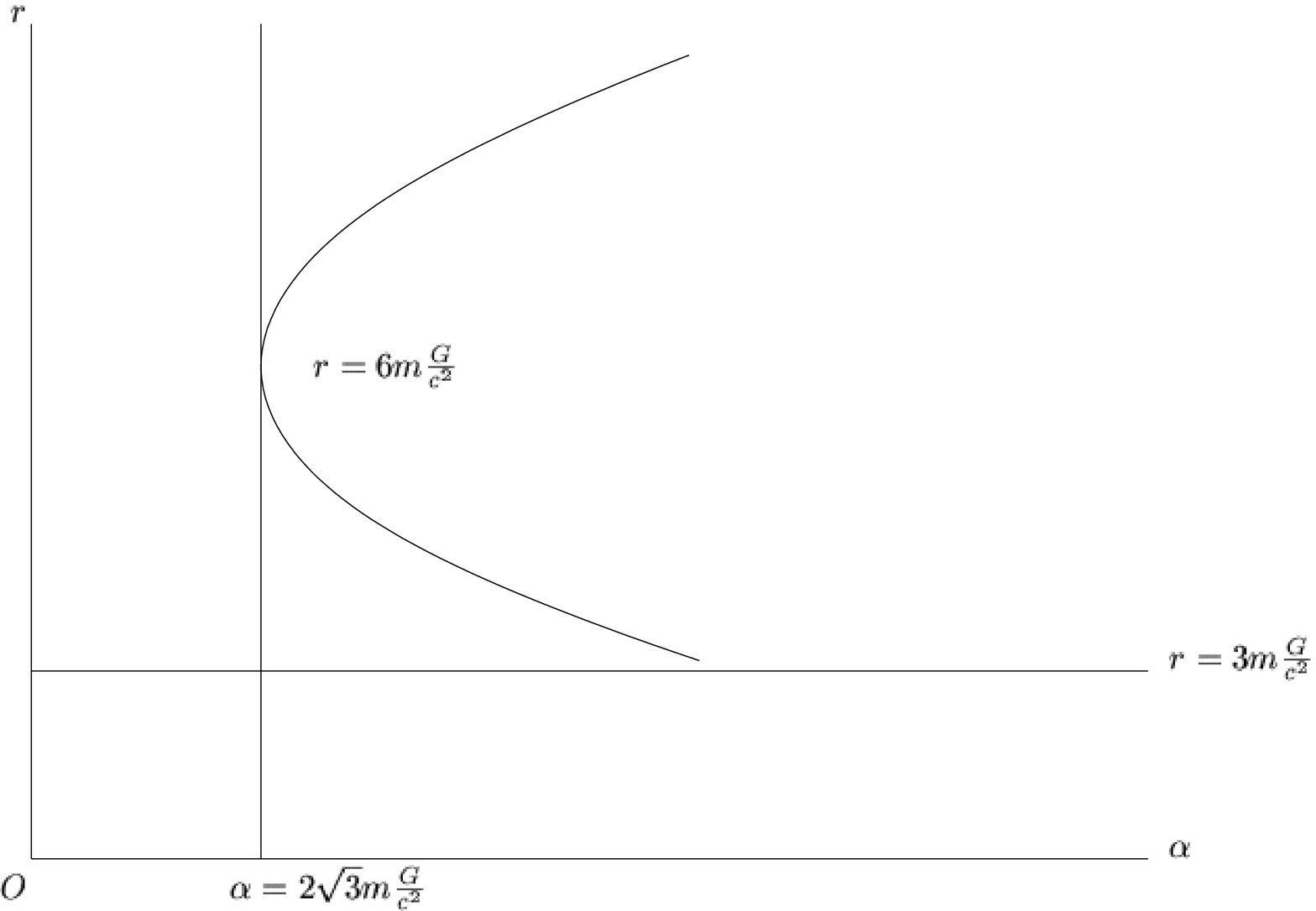}
\end{center}
\end{figure}

\clearpage 

\section{Appendix B: Figures}

\bigskip\bigskip\begin{center}Figure II.2\end{center}\bigskip\bigskip

\begin{figure}[h]
\begin{center}
\includegraphics[width=3.5in]{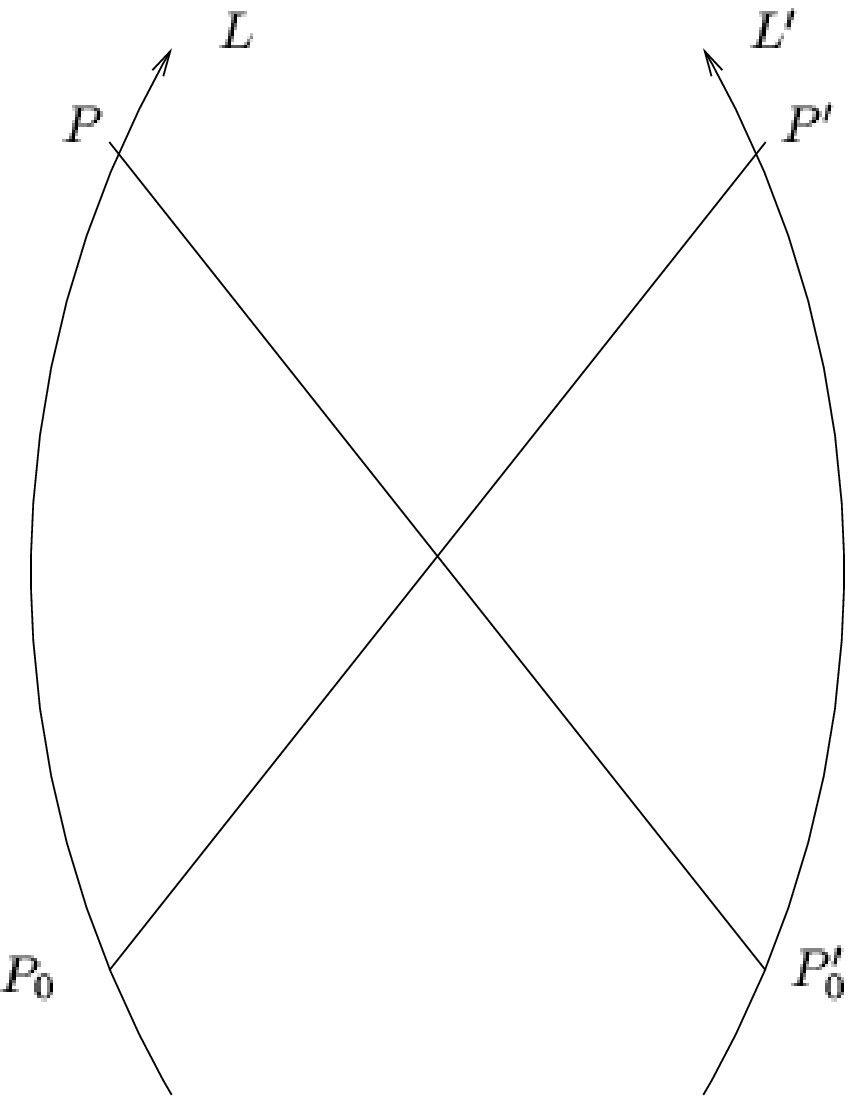}
\end{center}
\end{figure}

\clearpage 

\section{Appendix B: Figures}

\bigskip\bigskip\begin{center}Figure III.1\end{center}\bigskip\bigskip

\begin{figure}[h]
\begin{center}
\includegraphics[height=4.0in]
{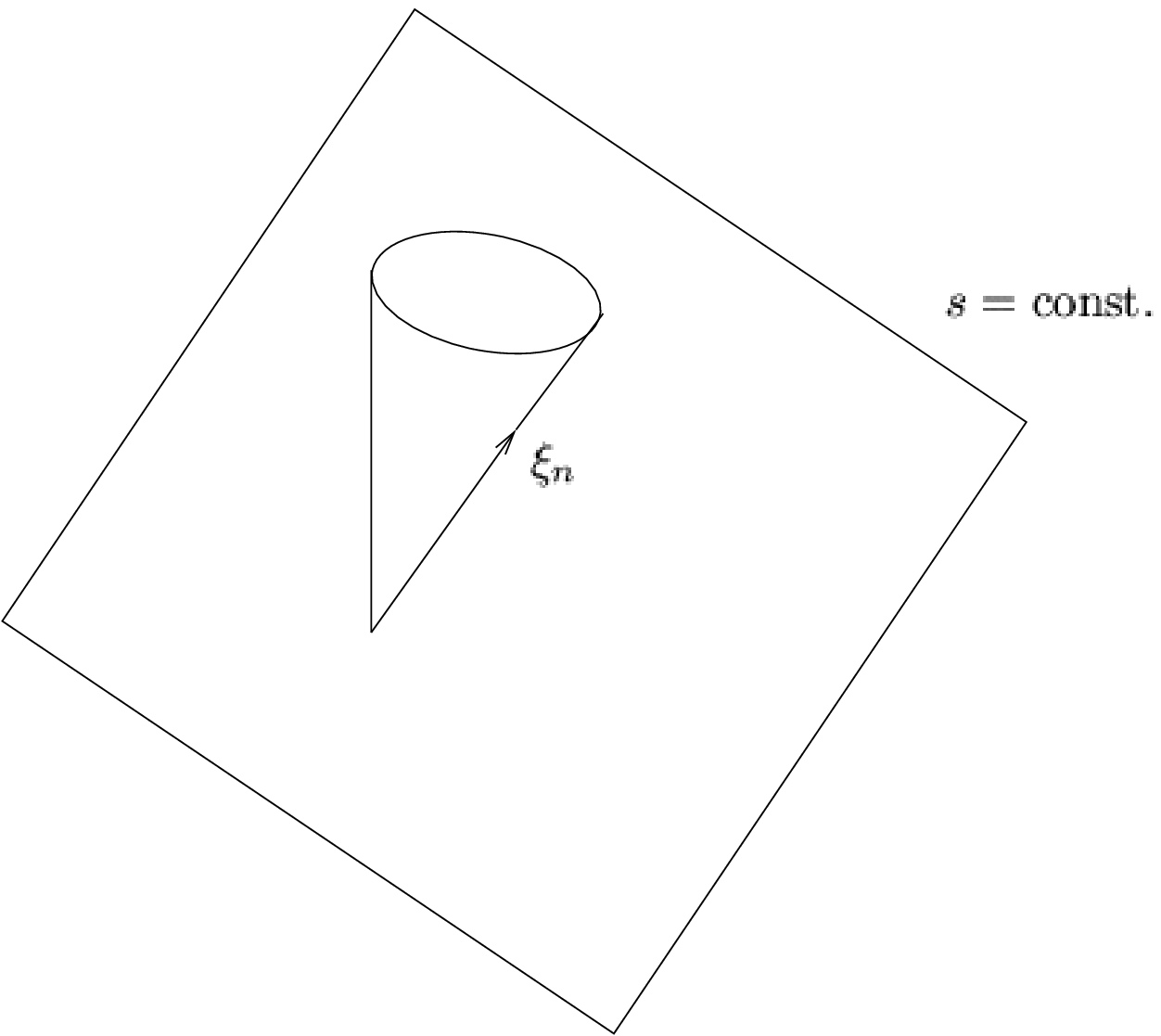}
\end{center}
\end{figure}

\end{document}